\documentclass[12pt,twoside,epsf]{article}
\usepackage{epsfig}
\usepackage{color}
\usepackage{bbold}
\usepackage{graphics}
\usepackage{booktabs}
\usepackage{amssymb}
\font\tenbf=cmbx9 %

\font\fourbf=cmbx14

\font\tenrm=cmr9

\font\tenit=cmti9 %

\font\sc=cmr12

\font\lrm=cmr9

\newcommand{\lle}{\mbox{$\langle$}}
\newcommand{\rle}{\mbox{$\rangle$}}

\newcommand{\bfsi}{\mbox{\boldmath$\sigma$}}

\newcommand{\bfep}{\mbox{\boldmath$\varepsilon$}}

\newcommand{\bfze}{\mbox{\boldmath$\zeta$}}

\newcommand{\bfcK}{\mbox{\boldmath$\cal K$}}
\newcommand{\bfcL}{\mbox{\boldmath$\cal L$}}

\newcommand{\bfa}{\mbox{\boldmath$\bf a$}}
\newcommand{\bfb}{\mbox{\boldmath$\bf b$}}

\newcommand{\bfg}{\mbox{\boldmath$\bf g$}}

\newcommand{\bfn}{\mbox{\boldmath$\bf n$}}

\newcommand{\bft}{\mbox{\boldmath$\bf t$}}
\newcommand{\bfs}{\mbox{\boldmath$\bf s$}}

\newcommand{\bfu}{\mbox{\boldmath$\bf u$}}

\newcommand{\bfx}{\mbox{\boldmath$\bf x$}}
\newcommand{\bfy}{\mbox{\boldmath$\bf y$}}
\newcommand{\bfz}{\mbox{\boldmath$\bf z$}}
\newcommand{\bfA}{\mbox{\boldmath$\bf A$}}

\newcommand{\bfC}{\mbox{\boldmath$\bf C$}}

\newcommand{\bfE}{\mbox{\boldmath$\bf E$}}
\newcommand{\bfG}{\mbox{\boldmath$\bf G$}}

\newcommand{\bfI}{\mbox{\boldmath$\bf I$}}
\newcommand{\bfJ}{\mbox{\boldmath$\bf J$}}
\newcommand{\bfL}{\mbox{\boldmath$\bf L$}}
\newcommand{\bfN}{\mbox{\boldmath$\bf N$}}
\newcommand{\bfP}{\mbox{\boldmath$\bf P$}}

\newcommand{\bfX}{\mbox{\boldmath$\bf X$}}

\newcommand{\bfU}{\mbox{\boldmath$\bf U$}}
\newcommand{\bfF}{\mbox{\boldmath$\bf F$}}
\newcommand{\bfR}{\mbox{\boldmath$\bf R$}}
\newcommand{\bfK}{\mbox{\boldmath$\bf K$}}
\newcommand{\bfM}{\mbox{\boldmath$\bf M$}}
\newcommand{\bfT}{\mbox{\boldmath$\bf T$}}

\newcommand{\bfbD}{\mbox{$\mathbb{D}$}}

\newcommand{\bfdelta}{\mbox{\boldmath$\delta$}}
\newcommand{\bfdel}{\mbox{\boldmath$\delta$}}

\newcommand{\bfLa}{\mbox{\boldmath$\Lambda$}}
\newcommand{\bfcD}{\mbox{\boldmath$\cal D$}}

\newcommand{\bfcG}{\mbox{\boldmath$\cal G$}}

\newcommand{\bfcE}{\mbox{\boldmath$\cal E$}}
\newcommand{\bfcR}{\mbox{\boldmath$\cal R$}}
\newcommand{\bfcA}{\mbox{\boldmath$\cal A$}}
\newcommand{\bfcB}{\mbox{\boldmath$\cal B$}}
\newcommand{\bfcC}{\mbox{\boldmath$\cal C$}}
\newcommand{\bfcJ}{\mbox{\boldmath$\cal J$}}

\newcommand{\bfal}{\mbox{\boldmath$\alpha$}}
\newcommand{\bfbe}{\mbox{\boldmath$\beta$}}
\newcommand{\bfga}{\mbox{\boldmath$\gamma$}}

\newcommand{\bftau}{\mbox{\boldmath$\tau$}}

\newcommand{\bfthe}{\mbox{\boldmath$\vartheta$}}
\newcommand{\bfeta}{\mbox{\boldmath$\eta$}}

\newcommand{\BB}{\begin{equation}}
\newcommand{\EE}{\end{equation}}

\newcommand{\BBEQ}{\begin{eqnarray}}
\newcommand{\EEEQ}{\end{eqnarray}}

\begin{document}


\centerline{\fourbf Additive general integral equations in }

\centerline{\fourbf thermoelastic micromechanics of composites }



\medskip
\vspace{12pt}
\centerline{{\bf{ Valeriy A.
Buryachenko\footnote{\tenrm Address all correspondence to V. Buryachenko:
Micromechanics and Composites, Cincinnati, OH 45202, USA; Buryach@yahoo.com}}}}

\vspace{4pt}
\centerline {\it Micromechanics and Composites, Cincinnati, OH 45202, USA}






\centerline{\vrule height 0.003 in width 16.0cm} \noindent
\begin{abstract}
This work presents an enhanced Computational Analytical Micromechanics (CAM) framework for the analysis of linear thermoelastic composite materials (CMs) with random microstructure. The proposed approach is grounded in an exact Additive General Integral Equation (AGIE), specifically formulated for compactly supported loading, including both body forces and localized thermal changes (such as those from laser heating). New general integral equations (GIEs) for arbitrary mechanical and thermal loading are proposed. 
A unified iterative solution strategy is developed for the static AGIE, applicable to CMs with both perfectly and imperfectly bonded interfaces, where the compact support of loading is introduced as a new fundamental training parameter. Central to this methodology is a generalized Representative Volume Element (RVE) concept, which extends Hill’s classical definition. The resulting RVE is not predefined geometrically, but rather emerges from the characteristic scale of the localized loading, effectively reducing the analysis of an infinite, randomly heterogeneous medium to a finite, data-driven domain. This generalized RVE approach enables automatic exclusion of unrepresentative subsets of effective parameters, while inherently eliminating boundary effects, edge artifacts, and finite size limitations. Moreover, the AGIE-based CAM framework is naturally compatible with machine learning (ML) and neural network (NN) architectures, facilitating the construction of accurate and physically informed surrogate nonlocal operators.
\end{abstract}

\centerline{\vrule height 0.003 in width 16.0cm} \noindent


\section{Introduction }

The most widely used methods in analytical micromechanics for random-structured composite materials (CMs) are built upon a small set of fundamental concepts. Among these, the Effective Field Hypothesis (EFH), the General Integral Equation (GIE), and the Representative Volume Element (RVE) are particularly prominent. On the one hand, these concepts serve as the core components of many established micromechanical models. On the other hand, they also act as limitations, constraining the generality and applicability of these methods.
In this work, we demonstrate that these traditional assumptions, while historically central, can be relaxed or even entirely abandoned, thereby opening new avenues for the generalization and advancement of micromechanical modeling.

The effective field hypothesis (EFH), originally introduced by pioneers such as Poisson, Faraday, Mossotti, Clausius, Lorenz, and Maxwell (1824–1879, see for references 
\cite{{Buryachenko`2007}, {Buryachenko`2022}}
posits that inclusions located at points $\bfx \in v_i$ experience a local homogeneous field that differs from the remote applied (macroscopic) field. EFH, denoted as {\bf H1a}, has become the cornerstone of analytical micromechanics and underpins most classical modeling approaches.
For over 150 years, EFH has guided the development of various analytical models and significantly advanced the field in ways unmatched by any other theoretical construct. EFH plays a dual role in micromechanics: it serves as both a conceptual framework for constructing micromechanical models and as an approximation technique for solving the general integral equation (GIE).
The GIE is an exact relationship that links the random fields at a given point to those in its surroundings. The origins of GIE trace back to Rayleigh (1892)
\cite{Rayleigh`1892}. In the context of linear elasticity, a comprehensive historical review of the classical GIE—where EFH is implicitly assumed—was presented by Buryachenko 
\cite{Buryachenko`2007}. 
More recently, a new class of GIEs has been developed that form the foundation for micromechanics without relying on EFH, as shown in \cite{Buryachenko`2022}.

These new GIEs are formulated in operator form using perturbators introduced by single or multiple inclusions embedded in an infinite matrix under an arbitrary effective field. This framework—referred to as computational analytical micromechanics (CAM)—does not depend on Green’s functions or specific constitutive models, thereby offering broader applicability. CAM-based GIEs significantly enhance the accuracy of local field predictions and, in some cases, even correct the sign of field estimations inside inclusions (see \cite{Buryachenko`2022}).
Widely used methods such as the Effective Field Method (EFM) and the Mori–Tanaka Method (MTM) can be viewed as specific solution strategies for the GIE, even if the term "GIE" is not explicitly invoked in those approaches.

The Representative Volume Element (RVE) is fundamental for predicting the effective behavior of heterogeneous materials. An appropriately sized RVE captures the material's microstructural heterogeneity while minimizing artificial size and boundary effects. According to Hill \cite{Hill`1963}, the RVE assumes macroscopically homogeneous boundary conditions (BC), enabling a well-defined effective moduli tensor.
RVE response is typically evaluated through Direct Numerical Simulations (DNS) of Microstructural Volume Elements (MVEs) derived from synthetic models or imaging techniques like micro-CT (see e.g. \cite{{Echlin`et`2014}, {Konig`et`1991}, {Ohser`M`2000}}). Selecting the proper RVE size requires balancing statistical representativeness and scale separation, often expressed as $a \ll \Lambda \ll L$, where $a$ is the microstructural scale, $\Lambda$ is the field variation scale, and $L$ is the macroscopic length. Under such separation, classical homogenization yields reliable effective properties.
To ensure RVE validity, its response must be independent of boundary conditions. A standard approach involves convergence studies, identifying the smallest domain yielding stable effective properties.
The related concept of Statistically Equivalent RVE (SERVE) leverages image-based, data-driven modeling to replicate microstructures for simulation. Various methodologies for defining RVE and SERVE sizes are detailed in \cite{{Bargmann`et`2018},
{Francqueville`et`2019}, {Harper`et`2012}, {Kanit`et`2003}, {Matous`et`2017}, {Moumen`et`2021}, {Ostoja`et`2016}, {Sab`N`2005}}.

When the scale separation assumption breaks down, the assumption of statistically homogeneous fields no longer holds. As a result, stress and strain averages become nonlocally coupled through a tensorial kernel, necessitating the use of an effective elastic operator in integral form. These operators include (see classification in \cite{Maugin`2017}) strongly nonlocal approaches (strain and displacement-type, peridynamics by Silling \cite{Silling`2000}, and weakly nonlocal models, including strain-gradient or stress-gradient formulations. In this context, micromechanics provides essential tools to bridge scales governed by nonlocal constitutive relations.
In place of classical effective moduli (as introduced by Hill \cite{Hill`1963}), one must use effective nonlocal operators, typically expressed in integral or differential form. This shift leads to a generalized RVE concept that accommodates both random structure \cite{{Drugan`2000},{Drugan`2003},{Drugan`W`1996}} and periodic CMs \cite{{Ameen`et`2018}, {Kouznetsova `et`2004a},{Smyshlyaev`C`2000}}.
The generalized RVE is key to capturing nonlocal phenomena arising from inhomogeneous fields, inherent material nonlocality, and long-range inclusion interactions.
Traditionally, both classical and generalized RVEs depend on predefined boundary conditions and an assumed form of the nonlocal operator, which can limit their generality. These constraints can be eliminated by introducing body forces or temperature variations with compact support (which corresponds to laser heating of CMs)—a new class of loading scenarios. Such problems are akin to those in functionally graded materials (FGMs) and can be effectively addressed using the GIE-CAM approach (see \cite{Buryachenko`2022}).
In general, this requires solving the General Integral Equation (GIE) in the entire space. However, the problem is significantly simplified through the introduction of the Additive General Integral Equation (AGIE), specially designed for localized loading conditions. In AGIE, the free term corresponds to the deformation of a homogeneous infinite matrix and is defined explicitly, while the renormalization term present in the original GIE is omitted. This simplification allows the AGIE solution to be formulated as a streamlined version of the classical GIE solution.
A major advantage of AGIE is that it redefines the RVE: instead of a theoretically infinite domain (as in Hill's classical concept \cite{Hill`1963}), the RVE is now determined solely by the compact support of loading, and is defined as the region outside of which strains and stresses vanish. This offers a fundamentally new and physically grounded perspective on the RVE in the presence of nonlocal effects and localized excitations.

In classical elasticity theory, interfaces between constituent phases in composite materials (CMs) are typically assumed to be perfectly bonded, implying continuity of both displacement and traction vectors across the interface. However, in many practical cases, interfacial imperfections are present. The literature describes two primary classes of models to represent such imperfect interfaces in CMs (see, e.g., \cite{{Bennet`L`2019}, {Dvorak`B`1992a}, {Duan`K`2007}, {Ibach`1997}, {Gurtin`et`1998}}).
The first class, often referred to as displacement-discontinuity interface models, allows for jumps in displacement across the interface. These include the dislocation-like model, free sliding model, linear-spring interface model, and its nonlinear extensions, such as the cohesive zone model, where the interface is treated as a zero-thickness layer of nonlinear springs governed by a specified traction–displacement relationship. 
The second class, known as the coherent interface model, involves traction discontinuities across the interface while maintaining displacement continuity. In this case, additional interface stresses arise from atomistic effects such as variations in coordination numbers, bond lengths, bond angles, and charge distributions at the interface (e.g., \cite{{Ibach`1997}, {Gurtin`et`1998}}). These interfaces remain mechanically “attached” to both the matrix and the inclusion phase, without introducing displacement jumps.
Both types of imperfect interface models are naturally incorporated into the Additive General Integral Equation (AGIE) framework proposed in this study, enabling a more realistic representation of interfacial behavior in micromechanical analysis.

The development of effective nonlocal operator theory has been significantly advanced by the integration of machine learning (ML) and neural networks (NNs), offering enhanced flexibility and modeling capacity. Early contributions by Silling \cite{Silling`2020} (see also 
\cite{{You`et`2020}, {You`et`2024}}) illustrated how Direct Numerical Simulation (DNS) data could be leveraged to construct surrogate integral operators for modeling complex materials.
More recently, nonlocal neural operators have emerged as powerful tools for learning mappings between function spaces, as shown in works such as \cite{{Li`et`2003},{Lanthaler`et`2024}}. A wide variety of neural operator architectures have been developed—DeepONet, PCA-Net, Graph Neural Operators, Fourier Neural Operator (FNO), and Legendre Neural Operator (LNO)—each optimized for different classes of operator learning problems. Comparative studies and performance evaluations can be found in
\cite{{Lanthaler`et`2024},{Gosmani`et`2022},{Hu`et`2024},{Kumara`Y`2023}}.

In the context of nonlocal mechanics, the Peridynamic Neural Operator (PNO) \cite{Jafarzadeh`et`2024} and its generalization to heterogeneous media, HeteroPNO \cite{Jafarzadeh`et`2024b}, have been introduced as physics-aware neural frameworks to model nonlocal interactions in peridynamic formulations.
In parallel, Physics-Informed Neural Networks (PINNs) have gained popularity for embedding governing physical laws directly into the training process as soft constraints \cite{{Raissi`et`2019, Karniadakis`et`2021, Hu`et`2024}}. When combined with neural operator architectures \cite{{Faroughi`et`2024}, {Gosmani`et`2022}, {Wang`Y`2024}}, these hybrid approaches enable highly accurate and physically consistent models of complex, nonlinear, heterogeneous, and nonlocal materials, while maintaining strong generalization performance across different input domains.

Although machine learning (ML) and neural network (NN) techniques have significantly advanced material modeling, they often overlook essential micromechanical principles—including scale separation, boundary effects, and the definition of a Representative Volume Element (RVE)—which are critical for accurate predictions in both linear and nonlinear settings. To address this limitation, the proposed methodology introduces a new class of compressed datasets tailored for complex microstructures, grounded in a redefined RVE framework.
This novel RVE concept departs from traditional dependencies on the constitutive laws of individual phases or the explicit analytical forms of surrogate operators. Instead, it employs field concentration factors within each constituent phase to represent the microstructure in a physically meaningful way. The resulting datasets, informed by this micromechanically consistent RVE, are designed to be universally applicable across a wide range of ML and NN architectures for learning nonlocal surrogate operators.
By eliminating issues related to size effects, boundary layers, and edge artifacts, this RVE-based approach enhances the reliability and accuracy of surrogate model predictions across various material systems and loading conditions.

The structure of the paper is organized as follows. Section 2 introduces the fundamental equations and notations of elasticity theory, along with the formulation of body forces and thermal changes with compact support, and the statistical characterization of random structured CMs.
Section 3 addresses the solution for an infinite homogeneous matrix subjected to localized loading. It includes decompositions of material properties and field variables, and presents both the General Integral Equations (GIEs) and the Additive General Integral Equations (AGIEs). These formulations incorporate either statistically averaged fields or fields induced in the infinite matrix by BFCS 
and TCCS loading.
Section 4 extends the AGIE framework to composites with imperfectly bonded interfaces, while Section 5 focuses on iterative solutions of AGIEs for random heterogeneous materials.
In Section 6, the concept of compressed datasets and a redefined RVE is developed for random structure CMs under body forces and thermal changes with compact support. This section also introduces the construction of effective elastic moduli and a family of surrogate nonlocal operators, trained using localized loading conditions as input parameters.


\section{Preliminary}
\setcounter{equation}{0}
\renewcommand{\theequation}{2.\arabic{equation}}

\subsection{Basic equations}

We consider a linear elastic body occupying an open, simply connected, bounded domain $w\subset R^d$
with a smooth boundary $\Gamma_0$ and with an indicator function $W$ and space dimensionality
$d$ ($d=2$ and $d=3$ for 2-$D$ and 3-$D$ problems, respectively).
The domain $w$ {\color{black} with the boundary $\Gamma^0$} contains a homogeneous matrix $v^{(0)}$ and a periodic
{\color{black} field} $X=(v_i)$ of heterogeneity $v_i$ with the centers $\bfx_i$, indicator functions $V_i$ and bounded by the closed
smooth surfaces $\Gamma_i$ $(i=1,2,\ldots)$.
It is presumed that the heterogeneities can be grouped into phases $v^{(q)} \quad (q=1,2,\ldots,N)$ with identical mechanical and geometrical properties.
We first consider the local basic equations of thermoelasticity of composites
\BBEQ
\label{2.1}
\nabla\cdot\bfsi(\bfx)&=&-\bfb(\bfx), \\ 
\label{2.2}
\bfsi(\bfx)&=&\bfL(\bfx)\bfep(\bfx)+\bfal(\bfx), \ \ {\rm or}\ \ \
\bfep(\bfx)=\bfM(\bfx)\bfsi(\bfx)+\bfbe(\bfx), \\ 
\label{2.3}
\bfep(\bfx)&=&\nabla^s\bfu, \ \
\nabla\times\bfep(\bfx)\times\nabla={\bf 0}, 
\EEEQ
where $\otimes$ and
and $\times$ are the tensor and vector products, respectively,
$\nabla^s$ is the symmetric gradient operator,
$\nabla^s\bfu:=[\nabla {\otimes}{\bf u}+(\nabla{\otimes}{\bf u})^{\top}]/2$, and $(.)^{\top}$ denotes matrix transposition;
$\bfb$ is the body force.
${\bf L(x)}$ and ${\bf M(x) \equiv L(x)}^{-1}$ are the known phase
stiffness and compliance tensors.
$\bfbe(\bfx)$ and $\bfal(\bfx)=-\bfL(\bfx)\bfbe(\bfx)$ are second-order tensors of local eigenstrains and eigenstresses.
In particular, for isotropic
phases, the local stiffness tensor $\bfL(\bfx)$ is presented in
terms of the local bulk $k(\bfx)$ and shear $\mu(\bfx)$
moduli and:
\BB
\label{2.4}
\bfL(\bfx)=(dk,2\mu)\equiv dk(\bfx)\bfN_1+2\mu(\bfx)\bfN_2, \ \ \bfbe(\bfx)=\beta^{t}\theta\bfdel,
\EE
${\bf N}_1=\bfdelta\otimes\bfdelta/d, \ {\bf N}_2={\bf I-N}_1$ $(d=2\ {\rm or}\ 3$) whereas
$\bfdelta$ and $\bfI$ are the unit second-order and fourth-order tensors; $\theta=T-T_0$ denotes the temperature changes with respect to a reference temperature $T_0$ and $\beta^{t}$ is a thermal expansion.
For all material tensors $\bfg$ (e.g., $\bfL, \bfM,\bfbe,\bfal)$ the notation $\bfg_1(\bfx)\equiv \bfg(\bfx)-\bfg^{(0)}=\bfg^{(m)}_1(\bfx)$ $(\bfx\in v^{(m)},\ m=0,1$) is exploited.
The upper index $^{(m)}$ indicates the
components, and the lower index $i$ shows the individual
heterogeneities; $v^{(0)}=w\backslash v$, $ v\equiv \cup v_i,
\ V(\bfx)=\sum V_i(\bfx)$, and $V_i(\bfx)$ are the
indicator functions of $v_i$, equals 1 at
$\bfx\in v_i$ and 0 otherwise, $(i=1,2,\ldots)$.
Substitution of Eqs. (\ref{2.2}) and (\ref{2.3}) into Eq. (\ref{2.1}) leads to a representation of the equilibrium equation (\ref{2.1}) in the form
\BB
\label{2.5}
{\bfcL}(\bfu,\bfal)(\bfx)+\bfb(\bfx)={\bf 0},\ \ \ {\bfcL}(\bfu,\bfal)(\bfx):=\nabla[\bfL\nabla\bfu(\bfx)+\bfal(\bfx)],
\EE
where ${\bfcL}(\bfu,\bfal)(\bfx)$ is an elliptic differential operator of the second order.
Hereafter, for the contraction of calculations, we introduce the substitutions
\BB
(\bfep,\bfsi) \leftrightarrow \bfthe,\ \ (\bftau,\bfeta) \leftrightarrow \bfze,\ \ (\bfal,\bfbe) \leftrightarrow \bfga
\label{2.6}
\EE
indicating that for each triplet $\bfthe=\bfep,\ \bfze=\bftau, \ \bfga=\bfal$ and
$\bfthe=\bfsi,\ \bfze=\bfeta, \ \bfga=\bfbe$, the subsequent equations with the variables $\bfthe,\bfze$ and $\bfga$ are reduced to the corresponding equations for the strain $\bfep$ and $\bfsi$. The stress and strain polarization tensors (which are simply a
notation convenience)
\BB
\label{2.7}
\bftau(\bfx)=\bfL_1(\bfx)\bfep(\bfx)+\bfal_1(\bfx), \ \ \bfeta(\bfx)=\bfM_1(\bfx)\bfsi(\bfx)+\bfbe_1(\bfx),
\EE
($\bfeta(\bfx)=-\bfM^{(0)}\bftau(\bfx)$),
respectively, vanishing in the matrix $\bftau (\bfy) \equiv \bfeta(\bfy)\equiv{\bf 0}$ ($\bfy\in v^{(0)}$).

The body force density $\bfb(\bfx)$ is assumed to have compact support (called body force with compact support, BFCS), be self-equilibrated, and vanish outside a specified loading region $\bfcB^b:=b({\bf 0}, B^b)$:
\BB
\label{2.8}
\int \bfb(\bfx)d\bfx={\bf 0}, \ \ \ \bfb(\bfy)\equiv {\bf 0}\ \ {\rm for}\ \ \bfy\not\in b({\bf 0}, B^b):=\{\bfy| |\bfy|\leq B^b\},
\EE
where $b({\bf 0}, B^b)$ denotes a ball of radius $B^b$ centered at the origin $\bfx = \mathbf{0}$.
The temperature changes $\theta$ with compact support (TCCS) $\bfcB^{\theta}:=b({\bf 0}, B^{\theta})$ is presented as:
\BB
\label{2.9}
\theta(\bfx)\equiv {\bf 0}\ \ {\rm for}\ \ \bfy\not\in b({\bf 0}, B^{\theta}):=\{\bfy| |\bfy|\leq B^{\theta}\},
\EE
i.e. $\bfal(\bfx),\bfbe(\bfx)$ have the same compact support $\bfcB^{\theta}$.

We also consider remote homogeneous boundary conditions (called the
kinematic uniform boundary conditions (KUBC) and static uniform boundary conditions
conditions (SUBC), respectively)
\BBEQ
\label{2.10}
\bfu(\bfy)&=& \bfep^{w_{\Gamma}}\bfy, \ \forall\bfy\in \Gamma_{0u}={\Gamma}_0,\\
\label{2.11}
\bft(\bfy)&=&\bfsi^{w_{\Gamma}}\bfn(\bfy), \ \forall\bfy\in \Gamma_{0\sigma}={\Gamma}_0,
\EEEQ
correspond
to the analyses of the equations for either strain or stresses, respectively, which are formally similar to each other.

\subsection{Statistical description of random structure composites}

We consider a popular group of composites, called matrix composites, which consists of a continuous matrix phase reinforced by isolated inhomogeneities of various shapes.
Three material length scales (see, e.g., \cite{Torquato`2002}) are considered:
the macroscopic scale $L$, characterizing the extent of $w$, and the microscopic scale $a$, related with the
heterogeneities $v_i$. Moreover, one supposes that the applied field varies on a characteristic length scale $\Lambda$.
The limit of our interests for both the material
scales and the field one are either
\BBEQ
\label{2.12}
\!\!\!\!\!\!\!\!\!\!\!\!\!\!\!L\geq\Lambda\geq a^{\rm int}\geq a\ \ \ {\rm or} \ \ L\gg \Lambda\gg a^{\rm int}\geq a\gg l_{\delta},
\EEEQ
where the inequalities (\ref{2.12}$_2$) are called a scale separation hypotheses.
The inequalities (\ref{2.12}) correspond to the case of random structure CMs, where $a^{\rm int}$ stands for the scale of long-range interactions of inclusions (e.g. $a^{\rm int}=6a$).

The random quantities of the statistically homogeneous random fields (see, e.g. \cite{{Buryachenko`2022},{Malyarenko`O`2019}})
are described by a conditional
probability density $\varphi (v_i,{\bf x}_i \vert v_1,{\bf x}_1$
for finding a heterogeneity of type $i$ with the center $\bfx_i$ in the domain $v_i$, with the fixed heterogeneities $v_1$ centered at ${\bf x}_1$.
The notation $\varphi (v_i , {\bf x}_i\vert ;v_1,{\bf x}_1)$ denotes the case ${\bf x}_i\neq
{\bf x}_1$.
We have $\varphi(v_i, {\bf x}_i\vert ;v_1,{\bf x}_1)=0$ {\color{black} (since heterogeneities cannot overlap) for values of ${\bf x}_i$ placed inside the
some domain $ v^0_{1}$ called ``excluded volumes'',} where $v^0_{1}\supset v_1$
with indicator function $V^0_{1}$ is the ``excluded volumes'' of $\bfx_i$ with respect to $v_1$,
and $\varphi (v_i, {\bf x}_i\vert ;v_1,{\bf x}_1)\to \varphi(v_i, {\bf x}_i)$
as $\vert {\bf x}_i-{\bf x}_m\vert\to \infty$, $m=1,\ldots,n$ (since no long-range order is assumed).
$\varphi (v_i,{\bf x})$ is a number density, $n^{(k)}=n^{(k)}({\bf x})$ of component $v^{(k)}\ni v_i$
at the point ${\bf x}$ and $c^{(k)}=c^{(k)}({\bf x})$ is the concentration, i.e. volume fraction,
of the component $v_i\in v^{(k)}$ at the point ${\bf x}$:
$ c^{(k)}({\bf x})=\langle V^{(k)}\rangle ({\bf x})=\overline v_in^{(k)}({\bf x}),
\ \overline v_i={\rm mes} v_i\ \ (k=1,2,\ldots,N;\ i=1,2,\ldots),\quad
c^{(0)}({\bf x})=1-\langle V\rangle ({\bf x}).$
{\color{black} Additionally to the average $\langle (.)\rangle ({\bf x})$,} the notation
$\langle (.)\vert; v_1,{\bf x}_1\rangle ({\bf x})$
will be used for the conditional average taken
for the ensemble of a statistically homogeneous
{\color{black} set} $X=(v_i)$ at the point ${\bf x}$,
on the condition that there is heterogeneity at
the point ${\bf x}_1$ and
${\bf x}_i\neq\bfx_1$.
The notation $\langle (.)\vert; v_1,{\bf x}_1\rangle ({\bf x})$
is exploited for the additional condition ${\bf x}\notin v_1$. In a general case
of {\it statistically inhomogeneous} media with the homogeneous matrix (e.g., for so
called {\it Functionally Graded Materials}, FGM), the conditional probability density is not invariant with respect to
translation, that is, the microstructure functions depend on their absolute positions
\cite{Buryachenko`2022}:
\BBEQ
\!\!\!\!\!\!\!\!\!\!\!\!\!\!\varphi (v_i , {\bf x}_i\vert ;v_1,{\bf x}_1)\!\!&=&\!\!
\varphi (v_i , {\bf x}_i +\bfx)\vert ;v_1,{\bf x}_1+\bfx), \ \ \ \nonumber\\
\label{2.13}
\!\!\!\!\!\!\!\!\!\!\!\!\!\!\!\!\!\!\!\!\!\varphi (v_i , {\bf x}_i\vert ;v_1,{\bf x}_1)&\not=&
\varphi (v_i , {\bf x}_i +\bfx)\vert ;v_1,{\bf x}_1+\bfx),\ \ \
\EEEQ
for statistically homogeneous $(\forall\bfx\in R^d)$ and inhomogeneous $(\exists \bfx\in R^d)$ media, respectively.

Of course, any deterministic (e.g., periodic) field of inclusions $v_i$ with the centers $\bfx_{\alpha}\in\bfLa$ can be presented by
the probability density $\varphi (v_i,{\bf x}_i )$ and conditional probability density
$\varphi (v_i,{\bf x}_i \vert; v_j,{\bf x}_j)$ expressed through the $\delta$ functions ($\bfx_{\bf \alpha}\in \bfLa$)
\BBEQ
\varphi (v_i,{\bf x}_i )&=&\sum_{\bf \alpha}\delta (\bfx_i-\bfx_{\bf \alpha}), \nonumber\\
\label{2.14}
\varphi (v_i,{\bf x}_i\vert; v_j,\bfx_j)&=&\sum_{\bf \alpha}\delta (\bfx_i-\bfx_{\bf \alpha})-\delta(\bfx_i-\bfx_j).
\EEEQ

\section{Additive general integral equations (AGIE)}
\setcounter{equation}{0}
\renewcommand{\theequation}{3.\arabic{equation}}

\subsection{Infinite homogeneous matrix subjected to compact support loading}

A linear-elastic reference material is introduced, characterized as a homogeneous stiffness tensor $\bfL^{(0)}$. Consider the governing equation for an infinite homogeneous medium occupying $\mathbb{R}^d$ ($d = 1, 2, 3$), subject to the body force density $\bfb(\bfx)$ defined in Eq. (\ref{2.9}):
\BB
\label{3.1}
{\bfcL}^{(0)}(\bfu^{b(0)})(\bfx)+\bfb(\bfx)={\bf 0},
\EE
where ${\bfcL}^{(0)}$ denotes the elliptic operator associated with the homogeneous stiffness tensor $\bfL^{(0)}$.
This force distribution induces a displacement field given by:
\BB
\label{3.2}
\bfu^{b(0)}(\bfx)\equiv - ({\bfcL}^{(0)})^{-1} \bfb.
\EE
Alternatively, this displacement can be represented via the Green operator $\bfG^{(0)}(\bfx)$ corresponding to the Navier equation (\ref{2.5}) for the homogeneous reference tensor $\bfL^{(0)}$:
\BB
\label{3.3}
\bfu^{b(0)}(\bfx)= \int \bfG^{(0)}(\bfx-\bfy)\bfb(\bfy)~d\bfy.
\EE
Loosely speaking, the Green operator $\bfG^{(0)}(\bfx)$ may be interpreted as the inverse of the reference stiffness operator, characterizing the response of the infinite medium to localized force distributions.
The strains and stresses can be obtained from Eq. (\ref{3.3})
\BBEQ
\label{3.4}
\bfep^{b(0)}(\bfx)&=& \int \nabla^s\bfG^{(0)}(\bfx-\bfy)\bfb(\bfy)~d\bfy,\\
\label{3.5}
\bfsi^{b(0)}(\bfx)&=& \bfL^{(0)} \int \nabla^s\bfG^{(0)}(\bfx-\bfy)\bfb(\bfy)~d\bfy.
\EEEQ
We now proceed to the consideration of the infinite matrix with properties $\bfL^{(0)}$ and $\bfbe^{(0)}(\bfx)$ subjected to zero body force $\bfb(\bfx)\equiv {\bf 0}$ and temperature changes $\theta$ (\ref{2.9}). Then for the pairs $(\bfep,\bfsi) \leftrightarrow \bfthe$, and $ (\bfal,\bfbe) \leftrightarrow \bfga$, the field distribution is described as
\BB
\label{3.6}
\bfthe(\bfx)=\int\bfU^{(0)\gamma}(\bfx-\bfy)\bfga^{(0)}(\bfy)~d\bfy,
\EE
with $\bfU^{(0)\gamma}=\bfU^{(0)},{\bf \Gamma}^{(0)}$. The strain Green tensor $\bfU^{(0)}$ is the second derivative of the reference Green tensor $\bfG^{(0)}$, i.e.,
$U^{(0)}_{ijkl}(\bfx)
=\nabla_j\nabla_ lG^{(0)}_{(ij)(kl)}$
of order $O\big(\int |\bfx|^{1-d}d|\bfx|\big)$ as $|\bfx|\to\infty$ vanishing at infinity ($|\bfx|\to\infty$).
Here, the symmetrization in the lower indices (denoted by parentheses) ensures the tensor satisfies the symmetry requirements of elasticity. The stress Green tensor ${\bf \Gamma}^{(0)}=-\bfL^{(0)}(\bfI\delta(\bfx)+\bfU^{(0)}(\bfx)\bfL^{(0)})$.

\subsection {Material and field decompositions}

The decomposition of material parameters
\BBEQ
\label{3.7}
\bfL(\bfx)=\bfL^{(0)}+\bfL_1(\bfx) , \ \ \bfga(\bfx)=\bfga^{(0)}(\bfx)+\bfga_1(\bfx)
\EEEQ
is equivalent to `extraction' of the material properties of the matrix, while the jumps of the corresponding material properties vanish inside the matrix $v^{(0)}$. Owing to compact support of the temeperature change $\theta(\bfx)$ (\ref{2.9}), $\bfga^{(0)}(\bfx)\not \equiv $const.
A similar idea is used for the decomposition of the field parameters based on the `extraction' of a field vis-à-vis the mechanical loading of the matrix
\BBEQ
\label{3.8}
\!\!\!\!\!\!\!\!\!\!\!\!\!\!\!\bfep(\bfx)\!&=&\! \bfep^I(\bfx)\!+\!\bfep^{II}(\bfx), \nonumber\\
\bfep^I(\bfx)&=&\bfep^I_0(\bfx)\!+\!\bfep^I_1(\bfx), \
\bfep^{II}(\bfx)= \bfep^{II}_0(\bfx)\!+\!\bfep^{II}_1(\bfx),\\
\label{3.9}
\!\!\!\!\!\!\!\!\!\!\!\!\!\!\!\bfu(\bfx)\!&=&\! \bfu^I(\bfx)\!+\!\bfu^{II}(\bfx), \nonumber\\
\bfu^I(\bfx)&=&\bfu^I_0(\bfx)\!+\!\bfsi^I_1(\bfx),
\bfu^{II}(\bfx)= \bfu^{II}_0(\bfx)\!+\!\bfu^{II}_1(\bfx),\\
\label{3.10}
\!\!\!\!\!\!\!\!\!\!\!\!\!\!\!\bfsi(\bfx)\!&=&\! \bfsi^I(\bfx)\!+\!\bfsi^{II}(\bfx), \nonumber\\
\bfsi^I(\bfx)&=&\bfsi^I_0(\bfx)\!+\!\bfsi^I_1(\bfx),
\bfsi^{II}(\bfx)= \bfsi^{II}_0(\bfx)\!+\!\bfsi^{II}_1(\bfx),
\EEEQ
with the sources
\BBEQ
\label{3.11}
\bfb^I(\bfx)&=&\bfb(\bfx),\ \ {\bfga}^I({\bf x})={\bf 0}, \\
\label{3.12}
\bfb^{II}(\bfx)&=&{\bf 0},\ \ {\bfga}^{II}({\bf x})={\bfga}({\bf x}).
\EEEQ
The decomposition fields (\ref{3.8})-(\ref{3.10}) are very popular (see e.g. \cite{Buryachenko`2022}) and usually used for the homogeneous displacement loading conditions:
\BB
\label{3.13}
\bfu^I(\bfx)=\bfep^{w\Gamma}\cdot \bfx,\ \ \bfu^{II}(\bfx)={\bf 0},
\EE
$\bfx\in{ w}_{\Gamma}$, $\bfep^{w_{\Gamma}}=$const.
Instead of popular homogeneous loading (\ref{3.13}), we consider the loading by both a forcing term (\ref{2.8}) and thermal changes
(\ref{2.9}) with compact support, when the conditions (\ref{3.12}) are replaced by (\ref{3.8})-(\ref{3.10}).
The fields $\bfep^I_0(\bfx)$, $\bfu^I_0(\bfx)$, $\bfsi^I_0(\bfx)$ and
$\bfep^{II}_0(\bfx)$, $\bfu^{II}_0(\bfx)$, $\bfsi^{II}_0(\bfx)$ correspond to the field inside the homogeneous infinite matrix subjected to the loading
(\ref{2.8}) and (\ref{2.9}), respectively.

Let us consider one inclusion $v_i$ inside an infinite homogeneous matrix.
The field $\bfu^I$ is estimated from two equations ($\bfu^I_0=\bfu^{b(0)}(\bfx)$)
\BBEQ
\label{3.14}
{\bfcL}^{(0)}(\bfu^I_0)
(\bfx) &=& -\bfb(\bfx),\\
\label{3.15}
{\bfcL}(\bfu)
(\bfx)
&=& {\bfcL}^{(0)}(\bfu^I_0).
\EEEQ
Substituting these into the equilibrium equation (\ref{2.1}) yields:
\BB
\label{3.16}
\nabla\bfL^{(0)}\nabla \bfu_1^I(\bfx)=- \nabla\bfL_1(\bfx)\nabla\bfu(\bfx),
\EE
which leads to an implicit integral representation of the strain field
\BB
\label{3.17}
\bfep^I(\bfx) =\bfep^{b(0)}(\bfx)+\int\bfU^{(0)}(\bfx-\bfy)\bfL_1^I(\bfy)\bfep^I(\bfy)V_i(\bfy)~d\bfy.
\EE

For estimation of the recidual fields $\{(\cdot)\}^{II}$, instead of Eqs. (\ref{3.14}) and (\ref{3.15}), we need to consider the equations
($\bfep^{II}(\bfx)=\bfep^{\theta}(\bfx)$)
\BBEQ
{\bfcL}^{(0)}(\bfu^{II}_0,\bfal^{(0)})
(\bfx) &=& {\bf 0},\nonumber \\
\label{3.18}
{\bfcL}^{(0)} (\bfu_1^{II},{\bf 0}))&=&-\bfcL_1(\bfu^{II}, \bfal_1)(\bfx).
\EEEQ
The right-hand side of Eq. (\ref{3.18}) is formally equivalent to the presence of
a fictitious body force
\BB
\label{3.19}
\nabla\bfL^{(0)}\nabla \bfu_1^{II}(\bfx)=- \nabla(\bfL_1(\bfx)\nabla\bfu^{II}(\bfx)+\bfal_1(\bfx)),
\EE
that allows us to reduce Eq. (\ref{3.19}) to an integral equation
by the use of the Green’s tensor $\bfG^{(0)}$ for the displacement
\BB
\label{3.20}
\bfu^{II}_1(\bfx) =\bfG^{(0)}*\nabla(\bfL_1\bfep^{II}+\bfal_1).
\EE
Symmetrized derivation of Eq. (\ref{3.20}) with subsequent integration of the
integral obtained by parts leads to the equations for strains
\BB
\label{3.21}
\bfep^{II}(\bfx) =\bfep^{\theta}(\bfx)+\int\bfU^{(0)}(\bfx-\bfy)[\bfL_1(\bfy)\bfep^{II}(\bfy)+\bfal_1(\bfy)]V_i(\bfy)~d\bfy.
\EE
where it was used that $\nabla_{\bf y}=-\nabla_{\bf x}$. Obtaining Eq. (\ref{3.21}) is similar to analogous manipulations with a constant
$\bfal^{(0)}$ (when $\theta\equiv$const, see, e.g., \cite{Buryachenko`2007}), althogh Eq. (\ref{3.21}) was obtained for compact support temperature changes
(\ref{2.9}).

Summation of Eqs. (\ref{3.17}) and (\ref{3.21}) like Eq. (\ref{3.7}$_1$) leads to total representation for strains ($\bfx\in v_i$)
\BB
\label{3.22}
\bfep(\bfx) =\bfep^{b(0)}(\bfx) +\bfep^{\theta}(\bfx)+\int\bfU^{(0)}(\bfx-\bfy)[\bfL_1(\bfy)\bfep(\bfy)+\bfal_1(\bfy)]V_i(\bfy)~d\bfy,
\EE
which can be recast in terms of
stresses ($\bfx\in v_i$)
\BB
\label{3.23}
\bfsi(\bfx) =\bfsi^{b(0)}(\bfx) +\bfsi^{\theta}(\bfx)+\int{\bf \Gamma}^{(0)}(\bfx-\bfy)[\bfM_1(\bfy)\bfsi(\bfy)+\bfbe_1(\bfy)]V_i(\bfy)~d\bfy,
\EE
by the use of identities
\BB
\label{3.24}
\bfL_1(\bfep- \bfbe) =-\bfL^{(0)}\bfM_1\bfsi\ \
{\rm and} \ \ \bfep = [\bfM^{(0)}\bfsi+\bfbe^{(0)}]+
[\bfM_1\bfsi+\bfbe_1].
\EE

\subsection{Additive general integral equations}

Equations (\ref{3.23}) and (\ref{3.24}) were obtained for one fixed inclusion $v_i$ inside an infinite matrix. For the composite material (CM), the direct summations of all the surrounding inclusions
exerting on the fixed inclusion $v_i$ is described by the GIE (called the Additive GIE, AGIE, see for details \cite{Buryachenko`2025})
($\bfx\in v_i$)
\BBEQ
\label{3.25}
!\!\!\!\!\!\langle {\bfthe} \rle_i (\bfx)&=& \bfthe^{b(0)} ({\bf x})+\bfthe^{\theta}(\bfx)+
\int \!\!\bfcL_j^{\gamma}(\bfx-\bfx_j,\bfze) \varphi (v_j,{\bf x}_j\vert v_i,{\bf x}_i)d{\bf x}_j
\EEEQ
for the triplet (\ref{2.6}).
The functions with the deterministic fields $\bfthe^{b(0)} ({\bf x})$ and $\bfthe^{\theta} ({\bf x})$ produced by the body force $\bfb(\bfx)$ (\ref{2.8}) and $\theta(\bfx)$ (\ref{2.9}) with compact supports in the infinite homogeneous matrix.
The tensors
\BBEQ
\label{3.26}
{\bfthe} (\bfx)-\overline{\bfthe} (\bfx):=\bfcL_j^{\theta\zeta}(\bfx-\bfx_j, \bfze) =\int \bfU^{\gamma}(\bfx-\bfy) \bfze(\bfy)V_j(\bfy) ~d\bfy
\EEEQ
called the {\it perturbators} have a physical meaning of perturbations ${\bfthe} (\bfx)-\overline{\bfthe} (\bfx)$ produced by inclusion $v_j$ in the area $\bfx\in v_i$ at the action of the effective field $\overline{\bfthe} (\bfx)$.
The superindeces $^{\theta\zeta}$ of the perturbators $\bfcL^{\theta\zeta}$ correspond to the variables in the left-hand side $\bfthe$ and right-hand side $\bfze$, respectively.
The term {\it Additive} GIE is used similarly to {\it Additive} Manufacturing because the perturbations in Eq. (\ref{3.25}) are directly added without any renormalization terms exploited in GIE (see details in Subsection 3.4).

To the best of the authors’ knowledge, there are currently no academic studies that explicitly address inhomogeneous thermal loading in composite materials (CMs), despite the well-recognized practical significance of this issue, particularly in contexts such as laser-based heating during additive manufacturing \cite{Yang`et`2019} and military applications \cite{Isakari`et`2017}. In these scenarios, laser pulse irradiation induces a rapid temperature rise in the localized region of exposure, generating steep temperature gradients and resulting in elevated thermal stress concentrations.
Conventional analyses typically consider thermal stress estimation in a homogeneous half-space subjected to laser heating (e.g., \cite{Yilbas`2013}). However, when laser-based techniques are applied to CMs, the resulting thermal and stress fields exhibit spatial variations on a length scale comparable to that of the microstructural heterogeneities (e.g., particle sizes \cite{Yang`et`2019}). This observation underscores the necessity of incorporating nonlocal effects, as formulated, for instance, in Eqs. (\ref{3.25})—to accurately capture the thermomechanical response of CMs under localized thermal excitation.

Figure 1 illustrates symbolic representations of loading with compact support, including a deterministic self-equilibrated body force $\bfb(\bfx)$, thermal variation field $\theta(\bfx)$, and an inhomogeneous eigenstress distribution $\bfal^{(0)}(\bfx)$. Additionally, the eigenstress mismatch $\bfal_1(\bfx)$ represents the contribution from the random inclusion field.

\subsection{General integral equations}
A key concept in analytical micromechanics is the GIE, which precisely relates random fields at a given point to those in its surroundings.
Various forms of GIEs, ranked by increasing generality, are detailed in \cite{Buryachenko`2015} and reproduced in \cite{Buryachenko`2022} ($\bfx\in w$):
\vspace{-2mm}
\BBEQ
\label{3.27}
\!\!\!\!\!\!\!{\bfep}({\bf x}) &=& {\bfep}^{w \Gamma}
+\int{\bfU}(\bfx-\bfy)[\bfL_1(\bfy)\bfep(\bfy)+\bfal_1(\bfy)]~d{\bf y},\\
\label{3.28}
\!\!\!\!\!\!\!{\bfep}({\bf x})&=&\langle {\bfep}\rangle ({\bf x})
+\int{\bfU}(\bfx-\bfy)\Big\{[\bfL_1(\bfy)\bfep(\bfy)+\bfal_1(\bfy)]\nonumber\\
&-&\overline{\lle} [\bfL_1(\bfy)\bfep(\bfy)+\bfal_1(\bfy)]\rle(\bfy)]~d{\bf y},\\
\label{3.29}
\!\!\!\!\!\!{\bfep}({\bf x})
&=&
\langle {\bfep}\rangle ({\bf x})
+\int[{\bfU}(\bfx-\bfy)[\bfL_1(\bfy)\bfep(\bfy)+\bfal_1(\bfy)]
\nonumber\\
&-&\underline{\lle}{\bfU}(\bfx-\bfy)[\bfL_1(\bfy)\bfep(\bfy)+\bfal_1(\bfy)]\rle(\bfy)]
~d{\bf y},
\EEEQ
$\lle(\cdot)\rle$ and $\lle(\cdot)\rle(\bfx)$ are the statistical averages introduced in Subsection 2.4.

\vspace{1.mm} \noindent \hspace{30mm}
\parbox{8.8cm}{
\centering \epsfig{figure=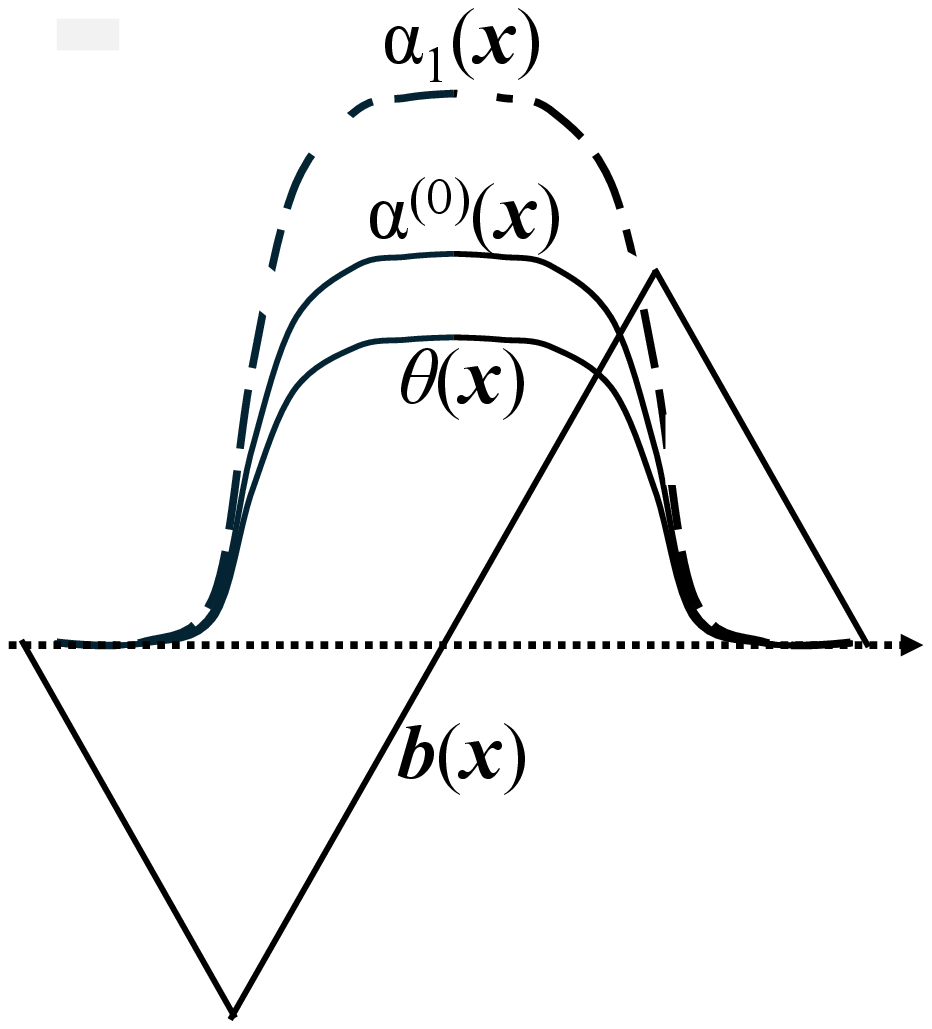, width=7.8cm}\\ \vspace{-50.mm}
\vspace{132.mm}
\vspace{-117.mm} \tenrm \baselineskip=8pt
{{ Fig. 1:} Deterministic $\bfb(\bfx),\ \theta(\bfx),\ \bfal^{(0)}(\bfx)$ and random $\bfal_1(\bfx)$ functions}}
\vspace{0.mm}

The historical evolution of the Generalized Integral Equation (GIE) from Eq. (\ref{3.27}) to Eq. (\ref{3.29}) -- highlighting key milestones such as those in \cite{{Rayleigh`1892},{Shermergor`1977},{Khoroshun`1978},{OBrian`1979}} -- is comprehensively reviewed in \cite{Buryachenko`2022}. Notably, the integral in Eq. (\ref{3.12}) lacks absolute convergence, in contrast to Eq. (\ref{3.13}), originally proposed by Rayleigh in \cite{Rayleigh`1892} under the assumptions of $\lle\bfep\rle(\bfx) \equiv {\rm const}$ and $\bfal_1(\bfy)\equiv{\bf 0}$. The latter avoids issues associated with the asymptotic behavior of the generalized Green operator $\bfU$ at infinity (decaying as $|\bfx - \bfy|^{1-d}$). Consequently, there is no requirement to define the shape or size of the integration domain $w$.
This also eliminates the need for regularization techniques, 
renormalization approaches, 
or auxiliary problems with mixed boundary conditions 
to address the divergence of the integral in Eq. (\ref{3.27}) at infinity (see for references and details \cite{Buryachenko`2022}). 
The transition of the statistical averaging operator from $\overline{\lle}$ in Eq. (\ref{3.28}) to $\underline{\lle}$ in Eq. (\ref{3.29}) signifies the birth of a second background of micromechanics -- known as Computational Analytical Micromechanics (CAM) \cite{Buryachenko`2022}. It was shown in \cite{Buryachenko`2022} that the exact Eq. (\ref{3.29}) simplifies to the approximate Eq. (\ref{3.28}) under the assumption of EFH validity.
CAM development represents a major advancement in the field since Rayleigh’s original intuitive introduction of the GIE framework in Eq. (\ref{3.28}) \cite{Rayleigh`1892}.

A centering procedure is applied to Eqs. (\ref{3.25}) by subtracting their statistical averages from both sides, leading to the more general Generalized Integral Equation (GIE) given in Eq. (\ref{3.29}). Unlike Eqs. (\ref{3.25}), which are only valid under the specific loading BFCS in Eq. (\ref{2.8}) and TCDC in Eq. (\ref{2.9}), the centered form in Eq. (\ref{3.29}) holds for arbitrary inhomogeneous loading fields $\langle \bfthe \rangle (\bfz)$. The key advantage of this centering is the emergence of the renormalizing term $\underline{\lle}{\bfU}(\bfx-\bfy)[\bfL_1(\bfy)\bfep(\bfy)+\bfal_1(\bfy)]\rle(\bfy)$ in Eq. (\ref{3.29}), which ensures the absolute convergence of the integral expressions.
Interestingly, the seemingly straightforward reduction of Eq. (\ref{3.26}) to Eq. (\ref{3.29})—and its subsequent simplification to Eq. (\ref{3.28})—offers an additional, less obvious benefit. Specifically, Eqs. (\ref{3.28}) and (\ref{3.29}) were obtained before when $\theta(\bfx) \equiv {\rm const.}$, such that for aligned, identical inclusions, the polarization term $\bfal_1(\bfy)$ is translation-invariant—even in the presence of a FGM as described by Eq. (\ref{2.13}$_2$). More remarkably, the same integral equations also valid when $\theta(\bfx) \not\equiv {\rm const.}$—as in the case of TCDC (\ref{2.9})—in which case $\bfal_1(\bfy)$ is no longer invariant
under translations, even for statistically homogeneous fields (\ref{2.13}$_1$) with aligned inclusions. To the best of the author’s knowledge, a generalization of Eqs. (\ref{3.28}) and (\ref{3.29}) to the nontrivial case $\theta(\bfx) \not\equiv {\rm const.}$ has never been undertaken.

Thus, two distinct developmental trajectories have emerged in the effort to rectify the limitations of the original Eq. (\ref{3.27}), each leading to two fundamentally new directions in the advancement of micromechanics. The first of these, initiated by Lord Rayleigh \cite{Rayleigh`1892}, entailed a critical modification of the integral term in Eq. (\ref{3.27}). This conceptual breakthrough laid the groundwork for more than a century of subsequent evolution, progressing from Eq. (\ref{3.28}) to its refined counterparts, Eqs. (\ref{3.29}). The culmination of this line of development in Eq. (\ref{3.29}) defines what may be considered the second background in micromechanics (see for details \cite{Buryachenko`2022}), and stands as the most significant advancement in the field since Rayleigh’s pioneering formulation \cite{Rayleigh`1892}.
In contrast, the second pathway retains the original integral structure of Eq. (\ref{3.27}) but introduces a critical conceptual shift by replacing the classical free term $\bfep^{w\Gamma}$ with the generalized term $\bfep^{b(0)}(\bfx)$ and $\bfep^{\theta}(\bfx)$, as specified in Eqs. (\ref{2.8}) and (\ref{2.9}), respectively. This replacement corresponds to a transition from traditional homogeneous boundary conditions (\ref{2.10}) to the field condition set BFCS and TCCS introduced in Eq. (\ref{2.8}) and
(\ref{2.9}). This alternative formulation, further elaborated in Eq. (\ref{3.25}), serves as the foundation for what is referred to in this work as the new philosophy of micromechanics (see Conclusion).
Far from representing a mere technical adjustment, this conceptual shift signifies a deep rethinking of the theoretical underpinnings of micromechanics—arguably the most transformative advance in the field since the classical contributions of Poisson, Faraday, Mossotti, Clausius, Lorenz, and Maxwell, 1824–1879; see \cite{Buryachenko`2022}), and the groundbreaking work of Lord Rayleigh
\cite{Rayleigh`1892}.

\section{AGIE for composites with imperfectly bounded interfaces}
\setcounter{equation}{0}
\renewcommand{\theequation}{4.\arabic{equation}}

The Averaged Generalized Integral Equation (AGIE) in Eq. (\ref{3.25}), along with the Generalized Integral Equations (GIEs) in Eqs. (\ref{3.28}) and (\ref{3.29}), are derived under the assumption that the composite material (CM) features perfectly bonded interfaces between its constituent phases. Under these conditions, both the displacement field and the traction components remain continuous across the interface boundaries. Specifically, the following continuity conditions hold:
\BB
\label{4.1}
[[\bfsi]]\cdot\bfn={\bf 0}, \ \ \ [[{\bf u}]]={\bf 0} 
\EE
on the interfacial boundary $\Gamma = \cup \Gamma_i$ ($i=1,\ldots$), which is assumed to be sufficiently smooth. Here, $\bfn$ denotes the outward unit normal vector on $\Gamma$ directed from the inclusion domain $v$ to the matrix domain $v^{(0)}$, and $[[(\cdot)]] = ({\rm out}) - ({\rm in})$ represents the standard jump operator.
The traction vector $\bft(\bfx) = \bfsi(\bfx)\bfn(\bfx)$ acting on a surface with normal vector $\bfn(\bfx)$ at point $\bfx$ can be expressed in terms of displacement as
$\bft(\bfx)=\hat{\bft}(\bfn,\nabla)\bfu(\bfx)+\bfal(\bfx)\bfn$,
where $\hat{t}_{ik}(\bfn, \nabla) = L{ijkl}n_j(\bfx)\frac{\partial}{\partial x_l}$ is the conormal derivative operator.

A brief overview, adapted from \cite{Buryachenko`2022} for further analysis, introduces imperfect interface models of the first and second kinds as slight generalizations of the perfect-bonded case.
In the first kind of imperfect interface model, the displacement components exhibit discontinuity across parts of the interface. Specifically, $[[{\bf u}(\bfs)]] \neq \mathbf{0}$ for points $\bfs \in \Gamma_i^u \subset \Gamma_i$, where $\Gamma_i^u$ denotes the subset of the interface with imperfect bonding. This discontinuity leads to a singular contribution to the strain field due to the differentiation of the discontinuous displacement field $\bfu(\bfs)$ on $\Gamma_i^u$. The strain tensor then takes the form:
\BB
\label{4.2} \bfep(\bfx) = \nabla^{S} \otimes \bfu(\bfx) + [[\bfu(\bfx)]]^{S} \otimes \bfn_i(\bfx)\delta_{\Gamma_i^u},
\EE
where $\bfn_i(\bfs)$ is the unit normal to the interface $\Gamma_i^u$, defined by the level-set equation $\Gamma_i^u(\bfx) = 0$, and $\delta_{\Gamma_i^u}$ is the Dirac delta distribution concentrated on $\Gamma_i^u$.

For a perfect interface, the last term in Eq. (\ref{4.2}) vanishes due to zero displacement jump. In contrast, a microcrack defined by $\Gamma_i = \Gamma_i^+ \cup \Gamma_i^-$ features zero traction on the crack faces: $\bfsi(\bfs^{\pm}) \cdot \bfn_i(\bfs^{\pm}) \equiv \mathbf{0}$.
Intermediate cases involve varying interactions between $\Gamma_i^+$ and $\Gamma_i^-$. A basic model treats displacement discontinuities as Somigliana dislocations \cite{Somigliana`1886}, with displacements $\bfb(\bfs)$ prescribed on the interface. Asaro \cite{Asaro`1975} showed that for ellipsoidal regions and polynomial $\bfb(\bfs)$, the resulting internal strain remains polynomial.
Grain boundary sliding, relevant even at room temperature, has been studied under frictionless interface conditions \cite{{Jasiuk`et`1987}, {Mura`F`1984}, {Mura`et`1985}}, and extended to graded inhomogeneities in \cite{Hatami-Marbini`S`2007}.
In the free sliding model, normal traction and displacement are continuous, while shear traction vanishes:
\BB \bfn \cdot [[\bfsi]] \cdot \bfn = 0,\quad [[\bfu]] \cdot \bfn = 0,\quad \bfP^s \cdot \bfsi = \mathbf{0}. \label{4.3}
\EE

Many researchers (e.g., \cite{{Dvorak`B`1992a},{Hashin`2002},{Huang`et`1993},{Zhong`M`1997}}) have studied the linear spring imperfect interface model (LSM), where displacement jumps across the interface are proportional to the local traction:
\BB
[[\bfsi]]\cdot \bfn = \mathbf{0}, \quad [[\bfu]] = \bfa \cdot \bfsi \cdot \bfn + \bfbe^{\rm int}, \label{4.4}
\EE with $\bfa$ containing compliance parameters $a_n$, $a_s$, and $a_t$ in normal and tangential directions, and $\bfbe^{\rm int}$ accounting for residual mismatches \cite{Duan`K`2007}.
These conditions decompose as:
\BB [[\bfsi]] \cdot \bfn = \mathbf{0}, \quad \bfP^s \cdot [[\bfu]] = a_s \bft^s + \bfP^s \cdot \bfbe^{\rm int}, \quad \bfP^n \cdot [[\bfu]] = a_n \bft^n + \bfP^n \cdot \bfbe^{\rm int}, \label{4.5}
\EE
where $\bft^s$ and $\bft^n$ are the shear and normal tractions. Perfect bonding corresponds to $a_n = a_s = 0$, while $a_n = 0$, $a_s \to \infty$ yields the free-sliding interface (\ref{4.3}). Finite values describe imperfect bonding.
Solutions for ellipsoidal inclusions with such interfaces (e.g., \cite{{Bennet`L`2019},{Dinzart`S`2017},{Lee`K`2019},{Othmani`et`2011}}) often rely on the thin-layer approximation.

The linear interface condition (\ref{4.4}) is extended in cohesive zone models (CZMs), where the continuous traction vector is a nonlinear function of the displacement jump (e.g., bilinear, trapezoidal, exponential, or polynomial laws). Linearizing this function near the origin recovers the spring-layer model. Originally introduced by Barenblatt \cite{Barenblatt`1962} for fracture mechanics, CZMs have since been widely adopted in micromechanics (e.g., \cite{{Othmani`et`2011}, {Needleman`1990},{Ortiz`P`1999},{Tan`et`2007},{Tvergaard`1990}}).

A second class of models, the Gurtin–Murdoch interface stress model (ISM) \cite{Gurtin`et`1998}, assumes displacement continuity, \
\BB [[\bfu(\bfs)]] = \mathbf{0}, \label{4.6}
\EE
and continuity of the surface-projected strain $\bfep^s = \bfP^s \bfep \bfP^s$, even though the bulk stress and strain fields can be discontinuous across the interface. ISM thus complements the spring-layer model as its dual formulation (see also \cite{Ibach`1997}, {Povstenko`1993}).

It should be noted that Eqs. (\ref{3.29}) are well-established for ideal interfaces $\Gamma_k$ (see Eq. (\ref{4.1})) and are presented here to illustrate the necessary corrections when discontinuities in displacement and/or stress fields exist across the interface $\Gamma_k$. In cases where $\bfu(\bfx)$ and $\bfsi(\bfx)$ exhibit discontinuities on the surfaces $\Gamma_k^u$ and $\Gamma_k^{\sigma}$, respectively—such that $\Gamma_k = \Gamma_k^u \cup \Gamma_k^{\sigma}$ and $\Gamma_k^u \cap \Gamma_k^{\sigma} = \emptyset$—Eqs. (\ref{3.16}) and (\ref{3.20}) must be interpreted in the sense of distributions, incorporating additional Dirac delta functions concentrated on the jump surfaces defined by $\Gamma_k^u(\bfx) = 0$ and $\Gamma_k^{\sigma}(\bfx) = 0$:
\BBEQ
\label{4.7}
\!\!\!\!\!\!\!\!\!\!\!\! \nabla\!\!\!\!&\cdot&\!\!\!\![\bfL^{(0)}(\bfx)(\nabla\
{\otimes}\bfu_1(\bfx))]=
-\nabla\cdot[\bfL_1(\bfx) (\nabla\
{\otimes}\bfu(\bfx))+\bfal_1(\bfx)]\nonumber \\
\!\!\!&-&\!\!\! [[\bfsi]]\cdot\bfn(\bfx)\delta_{\Gamma^{\sigma}_k}+\nabla\cdot\{\bfL_1^{(0)}(\bfx) ([[\bfu]]\
{\otimes}\bfn(\bfx))\}\delta_{\Gamma^{u}_k},
\EEEQ
where one typically assumes that either
\BB
\label{4.8}
\Gamma_k^u=\Gamma_k,\ \Gamma_k^{\sigma} =\emptyset \ \ \ {\rm or} \ \ \ \Gamma_k^{\sigma}=\Gamma_k,\ \Gamma_k^u =\emptyset.
\EE
By converting Eq. (\ref{4.7}) into its integral form and applying Gauss’ theorem, we obtain:
\BBEQ
\label{4.9}
\!\!\!\!\!\!\!\!\!\!\!\!\!\!\!\!\!\!\!\!\!\!\!\!\bfep(\bfx)&=&\bfep^{b(0)}+\bfep^{\theta}(\bfx)+\int_{v_k} \bfU(\bfx-\bfy)\bftau(\bfy)d\bfy
\nonumber\\
\!\!\!\!\!\!\!\!\!\!\!\!\!\!\!\!\!\!\!\!\!\!\!\!\!\!\!&-&\!\int_{\Gamma^u_k} \nabla \ \!^{^S}\!\!\!\!{\otimes}\bfT^{\top}(\bfx-\bfs)[[\bfu(\bfs)]]d\bfs-
\int_{\Gamma^{\sigma}_k}\nabla\ \!^{^S}\!\!\!\!{\otimes}\bfG(\bfx-\bfs)[[\bfsi(\bfs)]]\cdot\bfn(\bfs)d\bfs,
\EEEQ
Here, the surface integrals over $\Gamma_k^u$ and $\Gamma_k^{\sigma}$ in Eq. (\ref{4.9}) correspond, respectively, to those previously studied in \cite{{Asaro`1975} 
, {Sharma`G`2004}}, respectively. The tensor of ``fundamental traction" $\bfT$ on $\Gamma_k$ (called also Kuptadze tensor) assosiated with $\bfG^{(0)}$ is given by
\BBEQ
\label{4.10}
T_{ir}(\bfx,\bfs)=\bfL^{(0)}n_j^{\Gamma_k}(\bfs)\frac{\partial G^{(0)}_{pr}(\bfx-\bfs)}{\partial x_q}.
\EEEQ

To rewrite Eq. (\ref{4.9}) in terms of stresses, we use the identities (\ref{3.24}).
Substituting these into both sides of Eq. (\ref{4.9}) and contracting with ${\bf L}^{(0)}$ yields the corresponding stress-based integral equation
\BBEQ
\label{4.11}
\!\!\!\!\!\!\!\!\!\!\!\!\!\!\!\!\!\!\!\!\!\!\!\!\!\!\!\!\!\!&&\bfsi ({\bf x})=\bfsi^{b(0)}+\bfsi^{\theta}(\bfx)+\int_{v_k} {\bf \Gamma} (\bfx-\bfy)\bfeta({\bf y})
d{\bf y}
\nonumber\\
\!\!\!\!\!\!\!\!\!\!\!\!\!\!\!\!\!\!\!\!\!\!\!\!\!\!\!\!\!\!\!\!\!&-&\!\int_{\Gamma^u_k} \bfL^c\nabla \bfT^{\top}(\bfx-\bfs)[[\bfu(\bfs)]]d\bfs-
\int_{\Gamma^{\sigma}_k}\bfL^c\nabla\bfG(\bfx-\bfs)[[\bfsi(\bfs)]]\cdot\bfn(\bfs)d\bfs,
\EEEQ

The perturbators (\ref{3.26}) were introduced for ideal BC at the interface (\ref{4.1}). For imperfect interface (\ref{4.2})-(\ref{4.6}),
the perturbators of strain and stress, and displacement fields due to inserting a finite-sized heterogeneity at point $\bfx_k$
\BBEQ
\label{4.12}
\bfcL^{\epsilon}_k(\bfx-\bfx_k,\widehat{\bftau }) &\equiv& \bfep(\bfx)-\overline{\bfep}(\bfx),\\
\label{4.13}
\bfcL^{\sigma}_k(\bfx-\bfx_k,\widehat{\bftau }) &\equiv& \bfsi(\bfx)-\overline{\bfsi}(\bfx).
\EEEQ
These quantities represent the field perturbations caused by the heterogeneity and account for internal forces, eigenfields, and interfacial discontinuities. $\widehat{\bftau}$ and $\widehat{\bfeta}$ are the symbolic notations of dependance of
$\bfcL^{\epsilon}_k(\bfx-\bfx_k,\widehat{\bftau })$ and $\bfcL^{\sigma}_k(\bfx-\bfx_k,\widehat{\bfeta })$
on $\bftau(\bfx)$, $\bfeta(\bfx)$ ($\bfx\in v_k)$, respectively, and $[[\bfu(\bfx)]]$ ($\bfx\in \Gamma_k^u)$,
and $[[\bfsi(\bfx)]]$ ($\bfx\in \Gamma_k^{\sigma})$.
These perturbators can be computed using various numerical methods such as VIE, BEM, FEM, hybrid FEM-BEM, or multipole expansions—each with its own strengths and limitations, depending on the application (see \cite{{Buryachenko`2007}, {Liu`et`2011}}).

In particular, these perturbators can be expressed through the Green functions
\BBEQ
\!\!\!\!\!\!\!\!\!\!\!\!\!\!\!\!\!\!\!\!\!\!\!\!\bfcL^{\epsilon}_k(\bfx\!\!&-&\!\!\bfx_k,\widehat{\bftau })=\int_{v_k} \bfU^{(0)}(\bfx-\bfy)\bftau(\bfy)d\bfy-\int_{\Gamma^u_k} \nabla \ \!^{^S}\!\!\!\!{\otimes}\bfT^{\top}(\bfx-\bfs)[[\bfu(\bfs)]]d\bfs
\nonumber\\
\!\!\!&-&\!\!\!
\int_{\Gamma^{\sigma}_k}\nabla\ \!^{^S}\!\!\!\!{\otimes}\bfG^{(0)}(\bfx-\bfs)[[\bfsi(\bfs)]]\cdot\bfn(\bfs)d\bfs, \label{4.14}\\
\!\!\!\!\!\!\!\!\!\!\!\!\!\!\!\!\!\!\!\!\!\!\!\!\!\!\!\!\!\!\bfcL^{\sigma}_k(\bfx\!\!&-&\!\!\bfx_k,\widehat{\bfeta })=\int_{v_k} {\bf \Gamma}^{(0)} (\bfx-\bfy)\bfeta({\bf y})
d{\bf y}-\int_{\Gamma^u_k} \bfL^{(0)}\nabla \bfT^{\top}(\bfx-\bfs)[[\bfu(\bfs)]]d\bfs
\nonumber\\
\!\!\!&-&\!\!\!
\int_{\Gamma^{\sigma}_k}\bfL^{(0)}\nabla\bfG^{(0)}(\bfx-\bfs)[[\bfsi(\bfs)]]\cdot\bfn(\bfs)d\bfs,. \label{4.15}
\EEEQ

The perturbators $\bfcL^{\epsilon}_k(\bfx-\bfx_k,\widehat{\bftau })$ and $\bfcL^{\sigma}_k(\bfx-\bfx_k,\widehat{\bfeta })$ are determined by solving Eqs. (\ref{4.12})–(\ref{4.15}) along with the interface contact conditions on $\Gamma_k^u$ and $\Gamma_k^{\sigma}$. {\color{black}.These conditions may follow specific cases such as those described in (\ref{4.2})–(\ref{4.6}), where either displacement or traction discontinuities are enforced. However, for more realistic modeling, it is often necessary to adopt a generalized contact formulation that simultaneously accounts for both displacement and traction discontinuities—an approach increasingly explored in recent literature (see for references \cite{Buryachenko`2022}).
Regardless of the specific model adopted, the form of the perturbators given in Eqs. (\ref{4.12})–(\ref{4.15}) remains applicable for further analysis.

It is worth noting that Eqs. (\ref{4.12}) and (\ref{4.13}) are mathematically equivalent, which follows directly from the identities
\BB
\label{4.16}
\bfep=\bfM^{c}\bfsi+\bfbe^{c}+\bfeta,\ \ \ \bftau=-\bfL^{c}\bfeta. 
\EE
As a result, the strain and stress perturbators are directly related through the expression
\BB
\label{4.17}
\bfcL^{\sigma}_k(\bfx-\bfx_k,\widehat{\bfeta })=\bftau(\bfx)V_k(\bfx)+\bfL^{c}\bfcL^{\epsilon}_k(\bfx-\bfx_k,\widehat{\bftau }). 
\EE
This relationship holds regardless of the specific functional form used for the perturbators, confirming that both formulations—strain-based and stress-based—are interchangeable and consistent within the framework.

Thus, solutions (\ref{4.9}) and (\ref{4.11}) for a single inclusion $v_j$ in the infinite homogeneous matrix can be presented as
$\bfx\in R^d$)
\BBEQ
\label{4.18}
\bfep(\bfx)&=&\bfep^{b(0)}+\bfep^{\theta}(\bfx)+\bfcL^{\epsilon\tau}_k(\bfx-\bfx_k,\widehat{\bftau })\\
\label{4.19}
\bfsi ({\bf x})&=&\bfsi^{b(0)}+\bfsi^{\theta}(\bfx)+\bfcL^{\sigma\eta}_k(\bfx-\bfx_k,\widehat{\bfeta }).
\EEEQ

For composite materials (CMs), the cumulative influence of all surrounding inclusions acting on a fixed inclusion $\bfx\in v_i$ is captured by the Additive General Integral Equation, AGIE (see for details \cite{Buryachenko`2025}), specifically in the form similar to Eqs. (\ref{3.25}) ($\bfx\in v_i$)
\BBEQ
\label{4.20}
!\!\!\!\!\!\langle {\bfthe} \rle_i (\bfx)&=& \bfthe^{b(0)} ({\bf x})+\bfthe^{\theta}(\bfx)+
\int \!\!\bfcL_j^{\theta\zeta}(\bfx-\bfx_j, \widehat{\bfze}) \varphi (v_j,{\bf x}_j\vert v_i,{\bf x}_i)d{\bf x}_j
\EEEQ
for the triplet (\ref{2.6}).

The operators $\bfcL_j^{\theta\zeta}(\bfx - \bfx_j, \bfze)$ (\ref{3.26}) and $\bfcL_j^{\theta\zeta}(\bfx - \bfx_j, \widehat{\bfze})$ (\ref{4.12}) represent the fundamental micromechanical problem of a single inclusion embedded in an infinite homogeneous matrix. We now introduce another class of perturbators:
\BBEQ
\label{4.21}
\bfcL^{\theta}_k(\bfx-\bfx_k,\overline {\bfthe }) &\equiv& \bfthe(\bfx)-\overline{\bfthe}(\bfx),\\
\label{4.22}
\bfcL^{\theta}_k(\bfx-\bfx_k,\widehat{\overline {\bfthe }}) &\equiv& \bfthe(\bfx)-\overline{\bfthe}(\bfx),
\EEEQ
which present the solutions of problems (\ref{3.26}) and (\ref{4.12}), respectively.
Strictly speaking, the perturbators $\bfcL^{\theta\zeta}_i(\bfx-\bfx_k,\bfze)$
are just the notations of some problem that is destined to be solved in the equation
$\bfthe-\overline{\bfthe}=\bfcL^{\theta\zeta}_i(\bfx-\bfx_k,\bfze)$
while the perturbators $\bfcL^{\theta}_i(\bfx-\bfx_k ,\overline{\bfthe})$ are the solutions of this problem.
Since the subsequent treatment of $\bfcL^{\theta}_k(\bfx-\bfx_k,\overline {\bfthe }) $ and $\bfcL^{\theta}_k(\bfx-\bfx_k,\widehat{\overline {\bfthe }}) $ is identical, we will focus solely on the operator $\bfcL^{\theta}_k(\bfx - \bfx_k, \overline{\bfthe})$ in the discussion that follows.

The equations (\ref{4.21}) can be recast in other operator forms
\BB
\label{4.23}
\bfthe ({\bf x})={\bfcB}_i^{\theta}(\overline{\bfthe}) ({\bf x})+\bfC^{\theta}_i(\bfx),
\EE
where
\BBEQ
\label{4.24}
{\bfcB}^{\theta}_i(\overline{\bfthe}) ({\bf x})\!\!\!&=&\!\!\!
\bfcL^{\theta}_i(\bfx-\bfx_i,\overline{\bfthe })+\bfI\overline{\bfthe} ({\bf x})-\bfcL^{\theta}_i(\bfx-\bfx_i,{\bf 0}),\\
\bfC_i^{\theta}(\bfx)\!\!\!&=&\!\!\!\bfcL^{\theta}_i(\bfx-\bfx_i,{\bf 0}).
\EEEQ

\section{Solution of AGIE}
\setcounter{equation}{0}
\renewcommand{\theequation}{5.\arabic{equation}}

Consider two inclusions, $v_i$ and $v_j$, embedded in an infinite homogeneous matrix subject to an inhomogeneous field
$\widetilde{\bfthe}_{i,j}(\bfx)$ ($\bfthe=\bfep,\bfsi;\ \bfx\in R^d$). For points $\bfx\in v_i$, Eqs. (\ref{3.21}) and (\ref{3.22}) can be rewritten as:
\BBEQ
\label{5.1}
\!\!\!\!\!\!\!\!\!\!\!\!\!\!\!\!\!\!\!\!\!\!\!\!\bfthe(\bfx)\!\!\!&=&\!\!\!\overline{\bfthe}_i(\bfx)+\bfcL^{\theta}_i(\bfx-\bfx_i,\widehat{\bfthe}),\\
\label{5.2}
\!\!\!\!\!\!\!\!\!\!\!\!\!\!\!\!\!\!\!\!\!\!\!\!\overline{\bfthe}_i(\bfx)\!\!\!&=&\!\!\!
\bfcL^{\theta}_j(\bfx-\bfx_j,\widehat{\bfthe}).
\EEEQ
Combining these leads to ($\bfx\in v_i$):
\BB
\label{5.3}
\bfthe(\bfx)-\widetilde {\bfthe}_{i,j}(\bfx) -\bfcL^{\theta}_i(\bfx-\bfx_i,{\widetilde {\bfthe}}_{i,j})=
\bfcL^{\theta}_{i,j}(\bfx-\bfx_j,{\widetilde {\bfthe}}_{i,j})
\EE
where $\bfcL^{\theta}_{i,j}(\bfx-\bfx_j,{\widetilde {\bfthe}}_{i,j})$ represents the interaction-induced perturbation due to inclusion $v_j$, which can be computed using numerical methods (e.g., BEM, FEM, etc.), just like $\bfcL^{\theta}_i(\bfx-\bfx_i,{\overline {\bfthe}}_{i})$
The kernels of $\bfcL^{\theta}_i(\bfx-\bfx_i,{\widetilde {\bfthe}}_{i,j})$ and
$\bfcL^{\theta}_{i,j}(\bfx-\bfx_j,{\widetilde {\bfthe}}_{i,j})$ (\ref{5.3}) can be decomposed as:
\BB
\label{5.4}
\bfcL^{\theta}_{i,j}(\bfx,\bfy)=\bfcL^{\theta I}_{i,j}(\bfx,\bfy)+\bfcL^{\theta J}_{i,j}(\bfx,\bfy),\ \ \
\bfcL^{\theta K}_{i,j}(\bfx,\bfy)=\bfcL^{\theta}_{i,j}(\bfx,\bfy)V_k(\bfy),
\EE
where $K=I,J)$and summation follows Mura's \cite{Mura`1987} tensor notation.

From Eq. (\ref{5.3}), the field inside $v_i$ becomes:
\BB
\label{5.5}
\bfthe ({\bf x})-\bfcB^{\theta}_i(\widetilde{\bfthe}_{i,j})(\bfx)-\bfcC^{\theta}_i(\bfx)= \bfcB^{\theta}_{i,j}
(\widetilde {\bfthe}_{i,j})(\bfx)+
\bfC^{\theta}_{i,j}(\bfx),
\EE
with operator decomposition:
\BBEQ
\label{5.6}
\bfcL^{\theta}_{i,j}(\bfx-\bfx_j,{\widetilde {\bfthe}}_{i,j})=
\bfcB^{\theta}_{i,j}
(\widetilde {\bfthe}_{i,j})(\bfx)+
\bfC^{\theta}_{i,j}(\bfx).
\EEEQ

Now, consider replacing $v_j$ by a fictitious inclusion with properties $\bfcL_1(\bfy)\equiv {\bf 0}$ ($\bfy\in v_i$) and matrix
properties $\bfcL^{(0)}$. The fictitious eigenfield ($\bfga=\bfal,\bfbe$ and $\bfthe=\bfep,\bfsi$, respectively) are:
\BBEQ
\label{5.7}
\bfga^{\rm fict}_1 ({\bf x})\!\!\!&=&\!\!\! \bfcR^{\theta}_i(\widetilde{\bfthe}_{i,j})(\bfx)-\bfF^{\theta}_i(\bfx),
\EEEQ
so the constitutive law $v_j$ becomes
$\bfsi(\bfy)=\bfL^{(0)}\bfep(\bfy)+\bfbe_1^{\rm fict}(\bfy)$.
Using $\bfze (\bfy)= \bfga_1^{\rm fict}(\bfy)$, Eqs. (\ref{5.3}) and (\ref{5.5}) reduce to:
\BBEQ
\label{5.8}
\bfthe(\bfx)-\widetilde {\bfthe}_{i,j}(\bfx) -\bfcL^{\theta}_i(\bfx-\bfx_i,\widetilde {\bfthe}_{i,j})\!\!\!&=&\!\!\!
\bfcL^{\theta\infty}_{i,j}(\bfx-\bfx_j,\widetilde {\bfthe}_{i,j}),\\
\label{5.9}
\bfthe ({\bf x})-\bfcB^{\theta}_i(\widetilde{\bfthe}_{i,j})(\bfx)-\bfC^{\theta}_i(\bfx)
\!\!\!&=&\!\!\! \bfcB^{\theta\infty}_{i,j}
(\widetilde {\bfthe}_{i,j})(\bfx)+
\bfC^{\theta\infty}_{i,j}(\bfx),
\EEEQ
respectively, defining the new perturbator
$\bfcL^{\theta\infty}_{i,j}(\bfx-\bfx_j,{\widetilde {\bfthe}}_{i,j})$ ($\bfx\in v_i$) and the concentrators
$\bfcB^{\theta\infty}_{i,j}(\widetilde {\bfthe}_{i,j})(\bfx)$,
$\bfC^{\theta\infty}_{i,j}(\bfx)$
which can be estimated analogously to
$\bfcL^{\theta}_{i,j}(\bfx-\bfx_j,{\widetilde {\bfthe}}_{i,j})$ (\ref{5.8}), and
$\bfcB^{\theta\infty}_{i,j}(\widetilde {\bfthe}_{i,j})(\bfx)$,
$\bfC^{\theta\infty}_{i,j}(\bfx)$ (\ref{5.9}), respectively.
Thus, the tensors $\bfcB^{\theta}_{i}(\overline{\bfthe}_i)(\bfx)$ and $\bfC^{\theta}_{i}(\bfx)$ ($\bfx,\bfy\in v_i$)
define the field inside isolated $v_i$, while $\bfcB^{\theta}_{i,j}(\widetilde{\bfthe}_{i,j}(\bfx))$ and $\bfC^{\theta}_{i,j}(\bfx)$ ($\bfx\in v_i$) capture perturbations from interaction with $v_j$, and $\bfcB_{i,j}^{\theta\infty}(\widetilde{\bfthe}_{i,j})(\bfx)$ and $\bfC_{i,j}^{\theta\infty}(\bfx)$ represent effects due to $v_j$ under external loading alone.

In a similar manner, we can define the effective field perturbators $\bfcJ^{\theta }_{i,j}$,
$\bfT^{\gamma}_{i,j}(\bfx)$ and $\bfcJ^{\theta\infty }_{i,j}$,
$\bfT^{\gamma\infty}_{i,j}(\bfx)$
describing the perturbation of the effective field
$\overline{\bfthe}_i(\bfx)-\widetilde {\bfthe}_{i,j}(\bfx)$ introduced by both the heterogeneity $v_j$ and the fictitious inclusion
with the response operator $\bfcL^{(0)}$ and eigenfield $\bfga_1^{\rm fict}(\bfy)$ ($\bfy\in v_j,\ \bfx\in v_i,\
\bfz \in R^d$) (\ref{5.7}), (\ref{5.8}), respectively,
Similarly, we define the effective field perturbators:
\BBEQ
\label{5.10}
\overline{\bfthe}_i(\bfx)-\widetilde {\bfthe}_{i,j}(\bfx)\!\!\!&=&\!\!\!
\bfcJ_{i,j}^{\theta }(\widetilde{\bfthe}_{i,j})(\bfx)+\bfT^{\gamma}_{i,j}(\bfx),\\
\label{5.11}
\overline{\bfthe}_i(\bfz)-\widetilde {\bfthe}_{i,j}(\bfz)\!\!\!&=&\!\!\!
\bfcJ_{i,j}^{\theta\infty }(\widetilde{\bfthe}_{i,j})(\bfx)+\bfT^{\gamma\infty}_{i,j}(\bfx),
\EEEQ
and under constant fields $\widetilde{\bfthe}_{i,j}(\bfy)=$const., the operators reduce to:
\BBEQ
\label{5.12}
\bfJ_{i,j}^{\theta }(\bfx)= \lle\bfcJ_{i,j}^{\theta}(\bfx,\bfy)\rle_{(j)}, \ \ \
\bfJ_{j}^{\theta \infty}(\bfz)=\lle\bfcJ_{j}^{\theta \infty}(\bfz,\bfy)\rle_{(j)}
\EEEQ
($\bfy\in v_j,\ \bfx\in v_i,\ \bfz\in R^d$), where $\lle(\cdot)\rle_{(j)}$ denotes averaging over $v_j$.
Operators like $\bfcB^{\theta}_{i},\ \bfcB^{\theta}_{i,j}$, $\bfcJ^{\theta}_{i,j}$, $\bfcJ^{\theta \infty}_{i,j}$) with kernel $\bfcE(\bfx,\bfy)$ act on inhomogeneous functions $\bfg(\bfy)=\lle\widetilde{\bfthe}_{i,j}\rle(\bfy),\lle\overline{\bfthe}_{j}\rle(\bfy)$, as:
\BB
\label{5.13}
\bfcE(\bfg)(\bfx)=\int \bfcE(\bfx,\bfy)\bfg(\bfy)V_k(\bfy)d\bfy.
\EE
For constant $\bfg(\bfy)=\bfg\equiv$const. ($\bfy\in V_k$), this reduces to:
\BB
\label{5.14}
\bfcE(\bfg)(\bfx)=\bfE(\bfx)\bfg,\ \ \bfE(\bfx)=\int\bfcE(\bfx,\bfy)V_k(\bfy)d\bfy.
\EE
Finally, the relation between field and effective field perturbators is:
\BBEQ
\label{5.15}
\bfcJ_{i,j}^{\theta}(\widetilde{\bfsi}_{i,j})(\bfx)
=(\bfcB_i^{\theta})^{-1}(\bfcB^{\theta}_{i,j}(\widetilde{\bfsi}_{i,j}))(\bfx), \ \ \
\bfT^{\gamma}_{i,j}(\bfx)=(\bfcB_i^{\theta})^{-1}(\bfC^{\theta}_{i,j})(\bfx).\ \
\EEEQ

Fixing the inclusion $v_i$ within a composite material induces a random effective field $\overline{\bfthe}_i ({\bf x})$ as defined in Eq.
(\ref{3.26}), which for the case $n=1$ can be recast for ($\bfx\in v_i$) in the form:
\BBEQ
\label{5.16}
\langle \overline{\bfthe}_i\rangle(\bfx)\!\!\!&=& \bfthe^{b(0)} ({\bf x}) + \bfthe^{\theta} ({\bf x})
+\int \lle\bfcL^{\theta}_j(\bfx-\bfx_j,\overline{\bfthe}) \vert ; v_i,{\bf x}_i\rle_j
\nonumber\\
&\times&\varphi (v_j,{\bf x}_j\vert; v_i,{\bf x}_i)
d{\bf x}_j.
\EEEQ
In this expression, the conditional perturbator $\lle\bfcL^{\theta}_j(\bfx-\bfx_j,\overline{\bfthe}) \vert ; v_i,{\bf x}_i\rle_j$ can be represented using the explicit perturbator derived for the interaction between two heterogeneities subjected to the effective field $\widetilde{\bfthe}_{i,j}$, yielding:
\BBEQ
\label{5.17}
\langle \overline{\bfthe}_i\rangle(\bfx)\!\!\!&=&\!\!\!\bfthe^{b(0)} ({\bf x}) + \bfthe^{\theta} ({\bf x})
+\int [\bfcJ^{\theta}_{i,j}(\lle\widetilde{\bfthe}_{i,j}\rle)(\bfx)+
\bfT^{\gamma}_{i,j}(\bfx)
\nonumber\\
&\times& \varphi (v_j,{\bf x}_j\vert; v_i,{\bf x}_i)
d{\bf x}_j.
\EEEQ

Importantly, no assumptions are imposed on the microstructural topology, inclusion geometry, or the internal field distribution within the inclusions. The key computational advantage of Eq. (\ref{5.17}) lies in the fact that it does not require the use of the Green’s function or Eshelby-type solutions. This flexibility allows for the analysis of composites with arbitrary anisotropy (in matrix or inclusions), complex inclusion shapes, and diverse microstructural arrangements.
The derivation of Eq. (\ref{5.17}) relies solely on numerical solutions to canonical problems involving one or two inclusions embedded in an infinite medium subjected to uniform loading at infinity. Under additional assumptions, the involved tensors may be further expressed through Green’s functions, the classical Eshelby tensor (Eshelby, 1961), or the so-called exterior Eshelby tensor.
Equation (\ref{5.17}) is exact and incorporates two distinct types of statistically averaged effective fields, $\lle\overline{\bfthe}_{i}\rle(\bfx)$ and $\lle\widetilde{\bfthe}_{i,j}\rle(\bfx)$, which are solutions of an infinite hierarchy of integral equations analogous to Eq. (\ref{3.26}). In practice, this hierarchy is truncated, with Eq. (\ref{5.17}) forming the terminal level of closure, which will be discussed further in the next subsection under an appropriate closing assumption.

Generalized forms of the AGIE, Eqs. (\ref{5.16}) and (\ref{5.17}), were derived as extensions of the Green’s function-based AGIE given in Eq. (\ref{3.25}). In an analogous fashion, the Green’s function-based GIE (Eq. (\ref{3.29}), for (for $\theta(\bfx)\equiv 0$) was generalized to encompass broader formulations
\BBEQ
\label{5.18}
\langle \overline{\bfthe}_i\rangle(\bfx)\!\!\!&=&\!\!\!\langle \bfthe\rangle ({\bf x})
+\int \bigl\{\lle\bfcL^{\theta}_j(\bfx-\bfx_j,\overline{\bfthe}) \vert ; v_i,{\bf x}_i\rle_j
\varphi (v_j,{\bf x}_j\vert; v_i,{\bf x}_i)\nonumber\\
\!\!\!&-&\!\!\!\!\lle \bfcL_j^{\theta}(\bfx-\bfx_j,\overline{\bfthe}_j )\rle(\bfx_j)
\bigl\}d{\bf x}_j,\\
\label{5.19}
\langle \overline{\bfthe}_i\rangle(\bfx)\!\!\!&=&\!\!\!\langle \bfthe\rangle ({\bf x})
+\int \bigl\{[\bfcJ^{\theta}_{i,j}(\lle\widetilde{\bfthe}_{i,j}\rle)(\bfx)+
\bfT^{\gamma}_{i,j}(\bfx)]\varphi (v_j,{\bf x}_j\vert; v_i,{\bf x}_i)\nonumber\\
\!\!\!&-&\!\!\!\! [\bfcJ^{\theta\infty}_{i,j}(\lle\widetilde{\bfthe}_{i,j}\rle)(\bfx)+
\bfT^{\gamma\infty}_{i,j}(\bfx)]
\bigl\}d{\bf x}_j.
\EEEQ
It is evident that Eqs. (\ref{5.18}) and (\ref{5.19}) closely resemble Eqs. (\ref{5.16}) and (\ref{5.17}), respectively, though the latter are structurally simpler. In particular, the free terms $\bfthe^{b(0)} ({\bf x}) + \bfthe^{\theta} ({\bf x}) $ in Eqs. (\ref{5.16}) and (\ref{5.17}) represent deterministically precomputed fields as defined in Eqs. (\ref{3.17}) and (\ref{3.21}), whereas the free terms $\lle\bfthe\rle(\bfx)$ in Eqs. (\ref{5.18}) and (\ref{5.19}) denote a priori unknown statistical averages of the field $\bfthe(\bfx)$. Furthermore, the renormalization terms present in Eqs. (\ref{5.18}) and (\ref{5.19}) are absent in the simpler formulations (\ref{5.16}) and (\ref{5.17}).

Furthermore, Buryachenko \cite{Buryachenko`2014} examined various solution strategies for Eqs. (\ref{5.18}) and (\ref{5.19}) based on different assumptions, including the effective field hypothesis, closure approximations, and ellipsoidal symmetry of the composite microstructure. Importantly, all these approaches can be readily simplified to the straightforward solution of Eq. (\ref{5.17}). In particular,
iteration method used in \cite{Buryachenko`2014} leds to representations
\BBEQ
\label{5.20}
\lle\overline{\bfthe}\rle&=&\lim_{n\to\infty}\lle\overline{\bfthe}^{[n]}\rle_i(\bfx)={\bfbD}^{b \theta}(\bfthe^{b(0)},\bfthe^{\theta},\bfx),\\
\label{5.21}
\lle\bfthe\rle_i(\bfx)&=&\bfcB^{\theta}({\bfbD}^{b \theta}(\bfthe^{b(0)},\bfthe^{\theta},\bfx))+\bfcC^{\theta}(\bfx).
\EEEQ

Equation (\ref{5.20}) provides a way to compute the ensemble average of the field at the macroscopic point $\bfX$:
\BB
\label{5.22}
\langle {\bfthe}\rle(\bfX) :=c^{(0)}\lle\bfthe\rle^{(0)}(\bfX)+
c^{(1)}\lle\bfthe\rle^{(1)}(\bfX)
\EE
Here, the macroscopic statistical average $\lle\bfthe\rle(\bfX)$ is decomposed into two contributions: the average field in the matrix phase, $\lle\bfthe\rle^{(0)}(\bfX)$, and the average field within the inclusions, $\lle\bfthe\rle^{(1)}(\bfX)$.
To formalize this approach, we define an auxiliary domain $v^1_i(\bfX)$ characterized by the indicator function $V^1_i(\bfX)$ and bounded by the surface $\partial v^1_i(\bfX)$, which corresponds to the set of centers of ellipsoidal regions $v_q(\mathbf{0})$ translated around the fixed macroscopic point $\bfX$. The domain $v^1_i(\bfX)$ is constructed as the limit of a sequence $v^0_{ki} \to v^1_q(\bfX)$, where each ellipsoid $v_k$ progressively contracts toward the point $\bfX$.
Under this framework, the statistical expectation in Eq. (\ref{5.22}) can be reformulated in terms of a distributed volume force density acting within the region $v_q^l$ (with $\bfy \in v_q$), yielding the expression:
\BBEQ
\label{5.23}
\!\!\!\!\!\!\!\!\!\!\lle\bfthe\rle(\bfX)\!\!\!&=&\!\!\!
\bfthe^{b(0)}(\bfX)+\bfthe^{\theta}(\bfX)\nonumber\\
&+& c^{(0)}
\int \bfcL^{\theta}_q(\bfX-\bfx_q,{\bfbD}^{b \theta}(\bfthe^{b(0)},\bfthe^{\theta},\bfy))
\varphi (v_q,{\bf x}_q\vert; \bfX) d{\bf x}_q \nonumber\\
\!\!\!&+&\!\!\!
+\int_{v^1_i({\bf X})} n^{(1)}(\bfx_q)
\bfcL^{\theta}_q(\bfX-\bfx_q,{\bfbD}^{b \theta}(\bfthe^{b(0)},\bfthe^{\theta},\bfy))
d{\bf x}_q.
\EEEQ
In Eq. (\ref{5.23}), the first integral captures the matrix contribution (corresponding to the term $c^{(0)}\lle\bfthe\rle^{(0)}$ in Eq. (\ref{5.22})), while the second integral accounts for the inclusions’ contribution (associated with $c^{(1)}\lle\bfthe\rle^{(1)}$). Importantly, the conditional probability density $\varphi (v_q,{\bf x}_q\vert; \bfX)$ vanishes whenever the integration point $\bfx_q$ lies within the auxiliary inclusion domain $v^1_i(\bfX)$, i.e., $\varphi (v_q,{\bf x}_q\vert; \bfX) = 0$ for $\bfx_q \in v^1_i(\bfX)$.

The statistical field averages within the inclusions, $\lle\bfthe\rle_i(\bfx)$ for $\bfx \in v_i$ [see Eq. (\ref{5.21})], and the macroscopic statistical averages, $\lle\bfthe\rle(\bfX)$ [Eq. (\ref{5.23})], are derived for the general cases of the BFCS (\ref{2.8}) and the TCCS (\ref{2.9}). According to Eqs. (\ref{3.11}) and (\ref{3.12}), the components of the field decomposition—namely, $\lle\bfthe^I\rle_i(\bfx)$ and $\lle\bfthe^{II}\rle_i(\bfx)$ (for $\bfx \in v_i$), as well as their macroscopic counterparts $\lle\bfthe^I\rle(\bfX)$ and $\lle\bfthe^{II}\rle(\bfX)$—can also be obtained from Eqs. (\ref{5.21}) and (\ref{5.23}) by selecting appropriate source terms. Specifically, the type-I field components correspond to the case $\bfthe^{b(0)}(\bfX)\not\equiv \mathbf{0},\ \bfthe^{\theta}(\bfX) \equiv \mathbf{0}$
(or, that is tha same, $\bfb(\bfX)\not\equiv {\bf 0},\ \theta(\bfX)\equiv 0$), while the type-II components are obtained when $\bfthe^{b(0)} (\bfX)\equiv \mathbf{0},\ \bfthe^{\theta} (\bfX)\not\equiv \mathbf{0}$ (or $\bfb(\bfX)\equiv {\bf 0},\ \theta(\bfX)\not\equiv 0$).

For 2D locally elastic composite materials with non-canonical inclusion geometries, extensive numerical solutions of Eqs. (\ref{5.18}) and (\ref{5.19}) for the case 
$ {{\bfthe}}(\mathbf{x}) \equiv {\rm const} $ have been obtained using several computational techniques (FEA, VIEM, and the boundary integral equation method, BIEM), as reported in \cite{Buryachenko`2022}. The solutions of Eqs. (\ref{5.20})–(\ref{5.23}) for 1D peridynamic composites with ${\bfthe}^{b(0)}(\mathbf{x}) \not\equiv \mathbf{0}$ and ${\bfthe}^{\theta}(\mathbf{x}) \equiv \mathbf{0}$ were previously analyzed in \cite{Buryachenko`2023, Buryachenko`2023a}. Extending these results to the simpler governing Eqs. (\ref{5.16}) and (\ref{5.17}) for the general case ${\bfthe}^{b(0)}(\mathbf{x}) \not\equiv \mathbf{0}$ and ${\bfthe}^{\theta}(\mathbf{x}) \not\equiv \mathbf{0}$ is straightforward and may be carried out by interested readers.

\section{Prospective avenues for AGIE utilization }
\setcounter{equation}{0}
\renewcommand{\theequation}{6.\arabic{equation}}

\subsection{Compressed effective datasets for random structure CMs }

For arbitrary BFCS (\ref{2.8}) and TCCS (\ref{2.9}) loadings, we derived formal expressions for both the effective macroscopic and microscopic fields in a composite material (CM) with a random microstructure. The total fields $\{\cdot\}^T$ are decomposed into two components $\{\cdot\}^T=(\{\cdot\}^I, \{\cdot\}^{II})$, following the structure of Eqs. (\ref{3.8})–(\ref{3.10}).
We focus on the macroscopic strains
$\lle\bfep^T\rle(\bfx)$ and stress $\lle\bfsi^T\rle(\bfx)$ along with their microscopic counterparts $\lle\bfep^T\rle_i(\bfz,\bfx), $
and $\lle\bfsi^T\rle_i(\bfz,\bfx)$, where the subscript $i$ refers to a specific inclusion $v_i$.
To support numerical implementation and practical use, we construct an {\it effective dataset}
across multiple realizations of the applied compact support loading (\ref{2.8}) and (\ref{2.9}) ($\bfx\in R^d)$
\BBEQ
\label{6.1}
\!\!\!\!\!\!\!\!\!\!\!{\bfcD}^{I}&\!=\!&\{\bfcD^{I}_k\}_{k=1}^N, \ \ \ {\bfcD}^{I}_k=\{
\lle{\bfep}^I_k\rle(\bfb_k,0,\bfx), \lle\bfsi_k^I\rle(\bfb_k,0,\bfx),
\nonumber \\
\!\!\!\!\!\!\!\!\!\!\!&&\!\!\! \lle{\bfep}_{ik}^I\rle(\bfb_k,0,\bfz,\bfx),
\!\lle{\bfsi}_{ik}^I\rle(\bfb_k,0,\bfz,\bfx), \bfb_k(\bfx), 0\},\\
\label{6.2}
\!\!\!\!\!\!\!\!\!\!\!{\bfcD}^{II}&\!=\!&\{\bfcD^{II}_k\}_{k=1}^N, \ \ \ {\bfcD}^{II}_k=\{
\lle{\bfep}^{II}_k\rle({\bf 0},\theta_k,\bfx), \lle\bfsi_k^{II}\rle({\bf 0},\theta_k,\bfx),
\nonumber \\
\!\!\!\!\!\!\!\!\!\!\!&&\!\!\! \lle{\bfep}_{ik}^{II}\rle({\bf 0},\theta_k,\bfz,\bfx),
\!\lle{\bfsi}_{ik}^{II}\rle({\bf 0},\theta_k,\bfz,\bfx), {\bf 0},\theta_k(\bfx)\}.
\EEEQ
Each entry $\bfcD^{T}_k=\{\bfcD_k^I,\bfcD^{II}_k\}$ corresponds to a specific realization of the loading (\ref{2.8}) and (\ref{2.9}), providing both macro-scale and inclusion-level fields: the average strains $\lle{\bfep}_k^T\rle(\bfx)$ and stresses $\lle{\bfsi}^T_k\rle(\bfx)$
as well as the local inclusion fields $\lle\bfep_{ik}^T\rle(\bfz,\bfx):=\lle\bfep_{k}^T\rle_i(\bfz)$ and $\lle{\bfsi}_{ik}^T\rle(\bfz,\bfx):=\lle\bfsi^T_{k}\rle_i(\bfz)$.
The macroscopic coordinate $\bfx\in R^d$ spans the domain of the homogenized CM, while the local coordinate $\bfz\in v_i$ resides within the representative inclusion $v_i$.
This effective dataset
${\bfcD}^{T}$ forms a basis for data-driven modeling, enabling efficient retrieval and interpolation of effective response functions for new realizations (\ref{2.8}) and (\ref{2.9}) in a computationally efficient manner.

\subsection{ Representative volume element (EVE)}

The Representative Volume Element (RVE), introduced by Hill \cite{Hill`1963}, provides its rigorous foundation

\noindent {\bf Definition 1}. {\it Representative volume element (RVE) (a) is structurally entirely typical of the whole
mixture on average, and (b) contains a sufficient number of inclusions for the apparent overall moduli
to be effectively independent of the surface values of traction and displacement, so long as these
values are ‘macroscopically uniform’.... The contribution of this surface layer to any average can be
negligible by taking a sample large enough.}

A representative response can be estimated using direct numerical simulations (DNS) of microstructural volume elements (MVEs), either synthetically generated or experimentally extracted (e.g., via micro-computed tomography, CT). The homogenized stiffness properties, denoted by $\bfL^{\rm A}_{\rm KUBC}$ and $\bfM^{\rm A}_{\rm SUBC}$, are computed under kinematically uniform boundary conditions (KUBC) [Eq. (\ref{2.10})] and statically uniform boundary conditions (SUBC) [Eq. (\ref{2.11})], respectively. These properties generally differ; however, their discrepancy
\BB
\label{6.3}
\bfL^{\rm A}_{\rm KUBC}-(\bfM^{\rm A}_{\rm SUBC})^{-1}\to {\bf 0},
\EE
tends to diminish as the MVE size increases. This convergence behavior provides a basis for approximating the size of a representative volume element (RVE) and inferring the effective material moduli $\bfL^*$. The RVE concept inherently presumes the use of sufficiently large (ideally infinite) material domains (see, e.g., \cite{Ostoja`et`2016} and \cite{Buryachenko`2022}, p. 226).
It is important to note that while increasing the number of realizations in this intuitive
{\color {black} RVE} framework may yield statistical convergence in estimating $\bfL^*$ (i.e., the average sample response), it does not inherently eliminate edge-related scale effects.
The effective moduli $\bfL^*$ and field concentration factors $\bfA^*(\bfz)$ ($\bfz\in v_i$) are related by mutual coupling, see equations
\BB
\label{6.4}
^L\!\bfL^*=^L\!\bfL^{(0)}+^L\!\bfR^*, \ \ \ \lle\bfep\rle_i(\bfz)=^L\!\bfA^*(\bfz)\lle \bfep\rle.
\EE

The RVE concept has become so widely adopted that it is often treated as a heuristic rather than a strict mathematical definition. Engineers frequently rely on “intuitive” RVEs—volume elements that appear representative based on experience, especially when dealing with complex microstructures that resist formal homogenization. While such RVEs can offer quick estimates, they may be misleading when material behavior depends on fine microstructural details.
In practice, intuitive RVEs are typically constructed using two main approaches: (1) simulated random inclusion fields and (2) image-based models derived from micro-CT scans of real materials (see, e.g., \cite{{Konig`et`1991}, {Ohser`M`2000}, {Torquato`2002}}).

A pictorial interpretation of the RVE concept is illustrated in {\color {black} Fig. 2}, which shows a micro-CT scan of a heterogeneous material subjected to remote homogeneous loading, as defined by Eq. (\ref{2.10}). The sample is treated as an RVE under the assumption that its response under the applied boundary loading reflects the average behavior of the entire composite material. It helps convey the idea that if the sample size is appropriately chosen, the localized fluctuations due to microstructure can be averaged out, and the effective properties extracted from this RVE can be considered representative of the bulk material behavior.

\vspace{1.mm} \noindent \hspace{30mm}
\parbox{8.8cm}{`
\centering \epsfig{figure=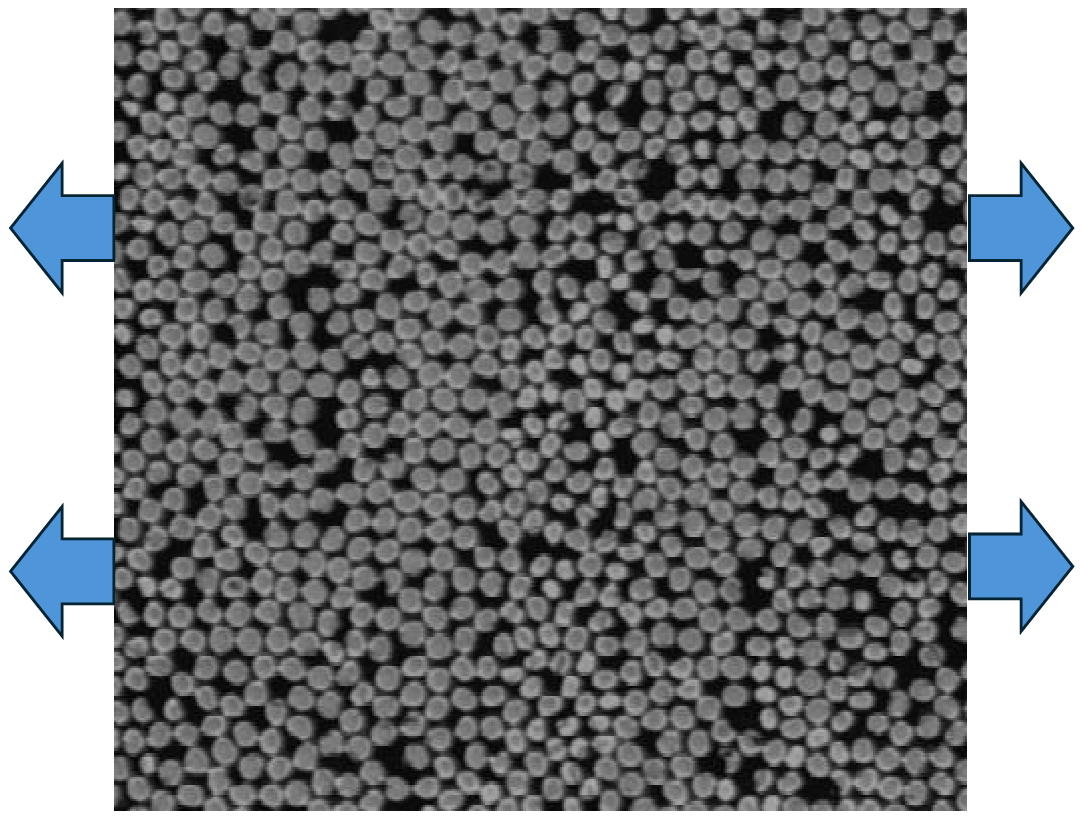, width=7.8cm}\\ \vspace{-50.mm}
\vspace{122.mm}
\vspace{-117.mm} \tenrm \baselineskip=8pt
{{ Fig. 2:} CT image with BC (\ref{2.4}) }}
\vspace{2.mm}

Reformulation and generalization of the classical definition \cite{Hill`1963}
enables one to formulate a flexible definition sufficient for our current interests in micromechanics with compact support loading (\ref{2.8}) and (\ref{2.9}) (rather than remote homogeneous loading (\ref{2.10}), see for details \cite{Buryachenko`2024b}):

\noindent {\bf Definition 2.} {\it RVE is structurally entirely typical of the whole CM area, which is sufficient for all apparent effective datasets $\bfcD^{I}$ (\ref{6.1}) and $\bfcD^{II}$ (\ref{6.2})
to be effectively stabilized outside RVE (e.g., vanishing of strains and stresses) in the infinite random structure CMs.}

{\color{black}
The RVE size is treated as a learned (or adjustable) parameter, selected so as to satisfy the tolerance criterion}
\BBEQ
\label{6.5}
{\color{black} |\bfsi(\bfx)|,|\bfep(\bfx)|<{\rm tol}\ \ {\rm for}\ \
\forall\bfx\in\overline{\rm RVR}:=R^d\setminus {\rm RVE}}
\EEEQ
{\color{black} which guarantees representativeness for homogenization. Moreover, since the applied body forces $\{\bfb_k(\bfx)\}_{k=1}^N$ have compact support, the infinite-domain problem reduces to a finite RVE, thereby eliminating finite-size and boundary (edge) effects.
}

This Definition 2 initiated in \cite{{Buryachenko`2023}, {Buryachenko`2023a}}, and formalized by Eqs. (\ref{6.1}) and (\ref{6.2}), differs fundamentally from Definitions 1. Notably, the idea of a “sample large enough” from Definition 1 is absent here. Instead, the heterogeneous medium is assumed to occupy the entire space $R^d$, whether statistically homogeneous or periodic. Additionally, unlike Definition 1, there is no mention of an ``effective moduli" (or ``effective nonlocal operator,” see \cite{Buryachenko`2022}). Rather, attention is focused on the domain $\overline {\rm RVE}=R^d\setminus {\rm RVE}$, where the effective datasets—$\bfcD^T$ —achieve stabilization.
Stabilization implies that all effective field parameters $\bfcD^T$ in the annular region between $|\bfx| = B^{\rm RVE}$ and $|\bfx| = B^{\rm RVE} + B^b/2$ remain consistent within a specified tolerance. In such cases, the region beyond $|\bfx| > B^{\rm RVE} + B^b/2$ may be excluded, allowing the infinite medium to be modeled by a finite sample. That is, a properly chosen RVE eliminates edge effects—effects which, in linear micromechanics, appear in boundary layers of thickness approximately $5a$ when $\lle\bfep\rle(\bfx)$ remains nonzero near the boundary (which is not our case; see p. 129 in \cite{Buryachenko`2007}).
{\color{black} Conversely, if $B^{\rm RVE}$ s too large for a given $\bfb_k(\bfx)$ (e.g.,
RVE$\not\subset \Omega_{00}^b$), residual boundary artifacts and finite-size effects contaminate the corresponding particular effective dataset $\bfcD_k^T$. Such datasets must therefore be excluded from further consideration. Consequently, the size B$B^b$ (or, equivalently, the gradient magnitude $|\nabla \bfb(\bfx)|$; see Eq. (2.8)) should be reduced.}

A key distinction must be emphasized between Definitions 1 and 2 of the RVE. Definition 1 pertains to an idealized, asymptotically large (infinite) domain, where effective material parameters are rigorously derived via a formal limiting process. In contrast, Definition 2 introduces a more practical notion—a finite-size RVE—that serves as an initial approximation, suitable for computational or experimental implementation and further refinement.
Despite this fundamental conceptual difference, the two definitions share an important technical similarity. In both cases, the effective properties are obtained through a limiting transition. Specifically, in Definition 1, the limit is taken in Eq. (\ref{6.3}) to derive the homogenized moduli (or operators, see \cite{Buryachenko`2022}) as defined in Eq. (\ref{6.4}$_1$). Analogously, in Definition 2, the process involves the convergence
$\bfcD \to \bfcD^{{\rm RVE}}$, where $\bfcD$ represents generalized effective dataset descriptors, extending the role of the quantities defined in Eq. (\ref{6.4}$_2$).
Notably, both definitions avoid the technical complications commonly encountered in the direct analysis of local stress singularities (in classical local models) or displacement jumps (in peridynamic models) that arise in specific realizations of heterogeneous media. These challenges are effectively bypassed by applying appropriate volume or statistical averaging procedures, which smooth out such local irregularities and ensure the robustness of the derived effective quantities.

With this modification, the domain of interest $\bfx \in w$ (refer to Fig. 3) is effectively reduced to the region $\bfx \in {\rm RVE}$ in accordance with Definition 2. The deterministic loading configuration within the original domain $\bfx \in w$ (as shown in Fig. 2) is replaced by the compactly supported loading described by Eqs. (\ref{2.8}) and (\ref{2.9}), now illustrated in Fig. 3. In this figure, the localized region of force application, $b(\bfx_i, B^b) \subset {\rm RVE}$, is explicitly depicted. This configuration emphasizes that the applied force is confined to a subregion within the RVE, thereby allowing the stabilization of the effective parameters $\bfcD^{\rm DNS}$ outside the RVE. The dataset $\bfcD^{\rm DNS}$ is generated from a series of realizations involving different CS loading
(\ref{2.8}) and (\ref{2.9}) and distinct microstructural patterns (e.g., obtained from various CT scans). Importantly, there are no constraints on the specific geometries of the regions $b(\bfx_i, B^b)$ [see Eq. (\ref{2.8})] or the RVE itself (as defined in Definition 2), nor on their relative sizes. The use of spherical shapes for these regions, as depicted in Fig. 3, is solely for illustrative simplicity and carries no intrinsic significance.

\vspace{1.mm} 
\hspace{15mm}
\parbox{8.8cm}{
\centering \epsfig{figure=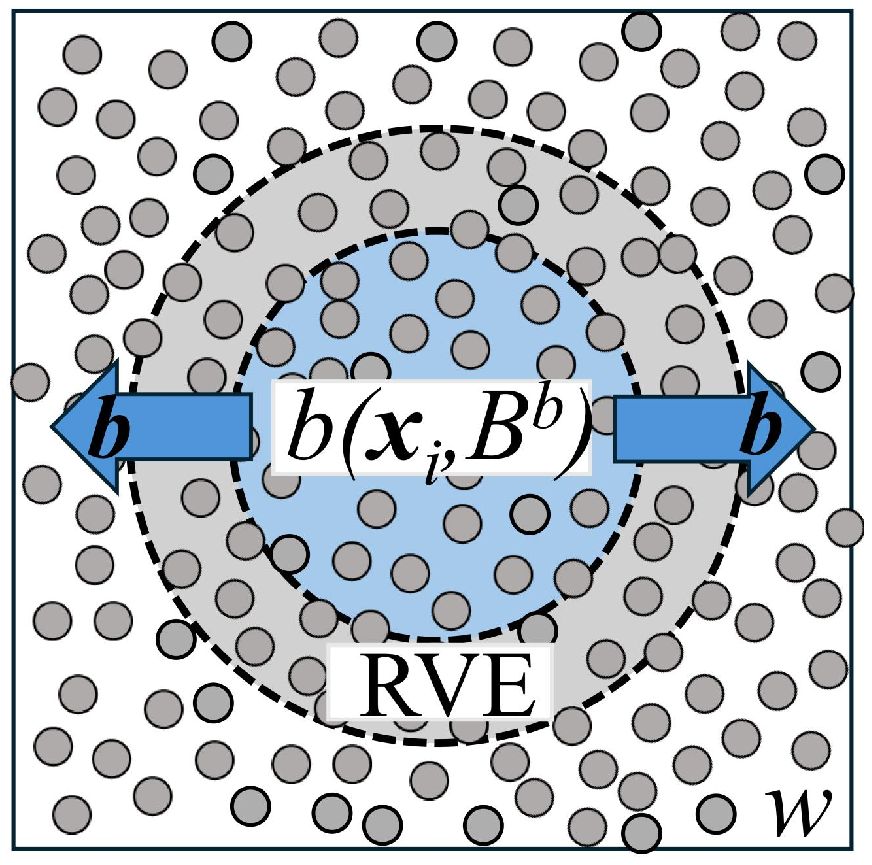, width=7.8cm}\\ \vspace{-50.mm}
\vspace{122.mm}
\vspace{-117.mm} \tenrm \baselineskip=8pt
{{ Fig. 3:} Scheme of CT image with $b({\bfx}_i,B^b)\subset{\rm RVE}\subset w$}}
\vspace{0.mm}

{\color {black} The significance of the RVE concept is drastically increased in peridynamic micromechanics, where there are three sorts of nonlocal effects generated by both
inhomogeneous applied fields $\bfb(\bfx)$, material nonlocality (horizon $l_{\delta}$), and interactions between inclusions.
A numerical interpretation of Fig. 3 is provided in \cite{{Buryachenko`2023},{Buryachenko`2023a}} for the case of a randomly inhomogeneous bar
under a self-equilibrated body force condition (\ref{2.8}), where $\bfb(\bfx) = 0$ for
$|\bfx| > B^b$ and $\bfb(\bfx)=-\bfb(-\bfx)$ for $|\bfx| \leq B^b$; $c^b=0$.
In Fig. 1 for effective displacent $\lle\bfu\rle(\bfX)$ vs $x/B^b$, curve 4 corresponds to the case $l_{\delta}=0$ (engineering approach) whereas the parameters $l_{\delta}/a=1$ and $c^{(1)}=0.5$ are fixed for corves 1-3;
$ B^b/a$ takes values
of $0.25,\ 0.375,\ 0.5$ (shown in curves 1, 2, and 3, respectively).

Due to nonlocal effects, the range of long-range interactions is restricted not only by the condition $|\bfx|\leq B^b$, but by a larger domain $|\bfx|\leq a^{\rm l-r}$ $(a^{\rm l-r}\approx 3 B^b$). As a result, the original problem posed on the infinite domain $\bfx\in R^1$ is effectively reduced to a finite domain.
This reduction implies that, for the considered scale ratios $a/B^{b}/l_{\delta}$, the domain size $|\bfx|\leq a^{\rm l-r}\approx 3 B^b$ requires stabilization of the displacement fields, $\lle\bfu\rle(\bfX)\approx$const. (or, equivalently, vanishing strain as expressed in Eq. (\ref{6.5})), $|\bfx|>a^{\rm l-r}$.
Although the term ``RVE” is not explicitly used, the domain $|\bfx|\leq a^{\rm l-r}$ in fact serves as the RVE for $\lle\bfu\rle(\bfX)$. Importantly, the size of this RVE depends on the scale ratios $a/B^{b}/l_{\delta}$.

\vspace{-1.mm}
\hspace{15mm}
\parbox{8.0cm}{
\centering \epsfig{figure=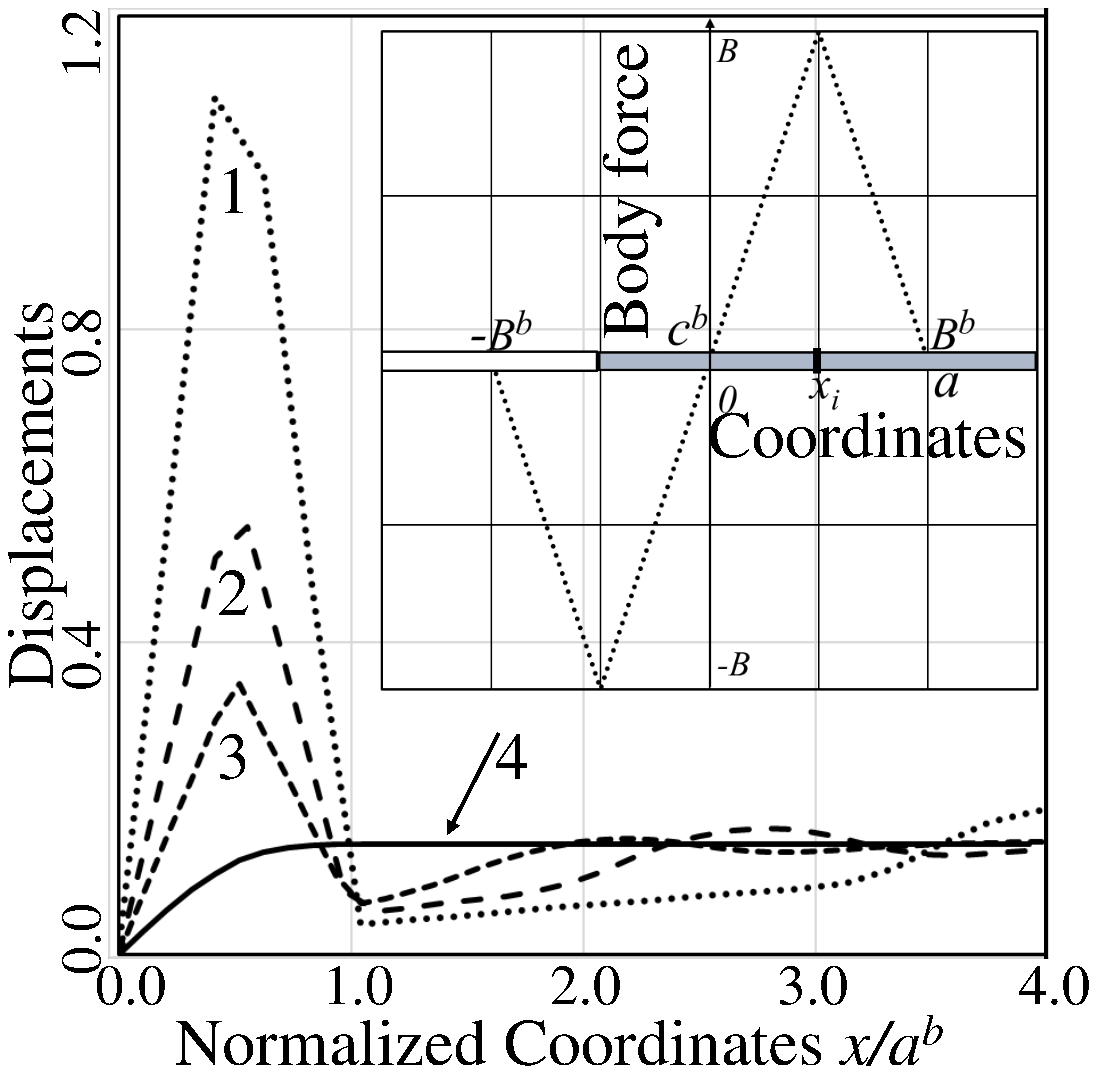, width=7.0cm} \vspace{-107.mm}
\vspace{80.mm}
\tenrm \baselineskip=8pt
{\hspace{1.mm}{\sc Fig. 4} Average displacement $\lle\bfu\rle(\bfx)$ vs $x/a^b$}}

}

A notable advancement in computational micromechanics is presented in the recent work by Silling {\it et al.} \cite{Silling`et`2024}, which proposes a novel coarse-graining strategy for two-dimensional random composites. The study examines a microstructure generated via Monte Carlo simulations in a square domain $w$, consisting of approximately 900 circular inclusions (Fig. 4a), with peridynamic behavior assigned to individual phases. Traditionally, such a configuration would suit classical RVE analysis under Definition 1 (see Fig. 2), using homogeneous periodic BC for homogenization.
In contrast, the authors employ a self-equilibrated body force $\bfb(\bfx)$ with compact support (\ref{2.8}) as in the approach of \cite{{Buryachenko`2023}, {Buryachenko`2023a}}. This satisfies the redefined RVE conditions of Definition 2, where ${\rm dist}(\partial {\rm RVE}, \partial w)\approx 10a=100\lambda$ with $\lambda$ denoting the lattice spacing. The effectiveness of this method is evident in the negligible strains in the violaceous boundary region near $\partial w$ (Fig. 5b), delineating $\overline{\mathrm{RVE}} = w \setminus \mathrm{RVE}$; {\color{black} this is a best illustration of vanishing condotion (\ref{6.5}))}.
While the term ``RVE” is not explicitly used in \cite{Silling`et`2024}, the multicolored interior region in Fig. 5b functionally embodies the generalized RVE concept. However, detailed field quantities like $\lle\bfep\rle_i(\bfz,\bfx),\ \lle\bfsi\rle_i(\bfz,\bfx)$ are left for future work.

Strikingly, the results for 1D cases in Fig. 4 from \cite{{Buryachenko`2023},{Buryachenko`2023a}} and Fig. 5b \cite{Silling`et`2024}—despite differing in dimensionality, microstructural character, and computational methodology—can all be interpreted via the generalized RVE concept (Definition 2). This unifying capability across approaches for both random \cite{Buryachenko`2023} and deterministic \cite{Silling`et`2024} structure models highlights the internal coherence, versatility, and innovative scope of the proposed RVE framework—an uncommon yet impactful development in micromechanics.

\subsection{Effective elastic moduli and surrogate operators }
\setcounter{equation}{0}
\renewcommand{\theequation}{7.\arabic{equation}}

An emerging and impactful direction in data-driven machine learning (ML) for modeling composite materials (CMs) was initiated by Silling \cite{Silling`2020} and further explored in \cite{You`et`2020, You`et`2024}. These works developed surrogate nonlocal operators for infinite media based on direct numerical simulations (DNS), focusing on a finite one-dimensional heterogeneous bar (featuring either periodic or random microstructures) subjected to two types of nonuniform loading: boundary-applied wave excitation and internal oscillating body forces. The corresponding dataset structure is defined as

\vspace{2.mm}
\hspace{-10mm} \parbox{12.8cm}{
\centering \epsfig{figure=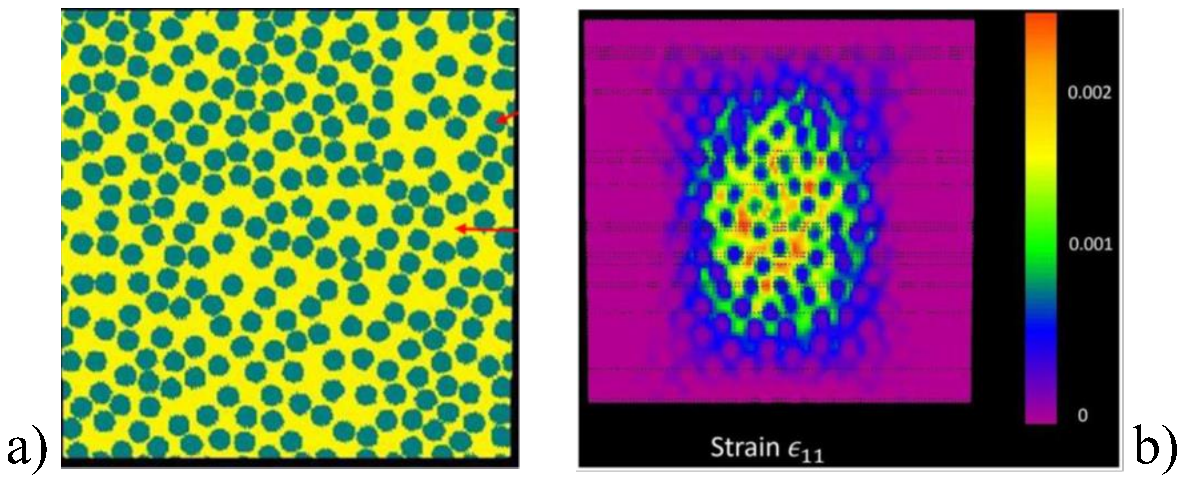, width=10.9cm}\\ \vspace{-22.mm}
\vspace{12.mm}
\vspace{-84.mm} \tenrm \baselineskip=8pt {\color{black}
{\hspace{5.mm}{\sc Fig. 5} a) Simulated structure in $w$. b) DNS of strains $\bfep(\bfx)$ in $w$}}
}\vspace{1.mm}

\BBEQ
\label{7.1}
\!\!\!\!\!\!\!\!\!\!\!{\bfcD}^{\rm DNS}=\{\bfcD^{\rm DNS}_k\}_{k=1}^N, \
\bfcD^{\rm DNS}_k=\{\bfu(\bfb_k,\bfx),\bfb_k(\bfx)\},
\EEEQ
where each sample represents a different realization of the body force $\bfb_k(\bfx)$, which does not necessarily satisfy the body-force compact support (BFCS) condition given in (\ref{2.8}).
By contrast, the present study (see \cite{Buryachenko`2023, Buryachenko`2023a}) extends this methodology to more general classes of random composite microstructures defined by Eq. (\ref{2.13}). Instead of relying on $\bfcD^{\rm DNS}$ (\ref{7.1}), this approach constructs effective datasets $\bfcD^{T}$ (see Eq. (\ref{6.1})) using BFCS $\bfb(\bfx)$ (\ref{2.8}) and TCCS (\ref{2.9}) loadings. These loading schemes serve both as excitation mechanisms and as well-controlled inputs for identifying nonlocal constitutive responses. The effective datasets are then compressed into reduced-order surrogate representations $\bfcD^{T}$, offering substantial reductions in computational effort while preserving key micromechanical characteristics.

Each surrogate dataset is employed to infer a corresponding nonlocal operator $\bfcL_{\rm p}$, formulated as a convolution-type integral operator:
\BBEQ
\!\!\!\!\!\!\!\!\!\!\!\!\!\!\bfcL_{\gamma}^{uT}[\lle{\bfu}^T_k\rle](\bfx) &=& {\bf \Gamma}^T(\bfx), \nonumber\\
\label{7.2}
\!\!\!\!\!\!\!\!\! \!\!\!\!\!\bfcL^{uT}_{\gamma}[\lle\bfu^T_k\rle](\bfx) &=&
\!\!\int \!\!\bfK^{uT}_{\gamma}(|\bfx-\bfy|) (\lle\bfu^{T}_k\rle(\bfy)-\lle\bfu^T_k\rle(\bfx))~d\bfy,
\EEEQ
where the subscript ${\gamma}=b,\sigma,u_i,\sigma_i$ designates one of four distinct operator types, corresponding to either strain, stress, or their localized variants. The quantity ${\bf \Gamma}^T_k(\bfx) := {{\bf \Gamma}^I_k(\bfx), {\bf \Gamma}^{II}_k(\bfx)}$ represents the averaged response fields—such as $\bfthe^{b(0)}$, $\langle \bfsi^I\rangle(\bfx)$, $\langle \bfep^I\rangle_i(\bfx)$, $\langle \bfsi^I\rangle_i(\bfx)$ for the mechanical case $\{\cdot\}^I$, Eq. (\ref{3.11})), or $\bfthe^{\theta}$, $\langle \bfsi^{II}\rangle(\bfx)$, $\langle \bfep^{II}\rangle_i(\bfx)$, $\langle \bfsi^{II}\rangle_i(\bfx)$ for the thermal case $\{\cdot\}^{II}$, Eq. (\ref{3.12})). These correspond to four types of learned surrogate operators:
$\bfcL^{uT}_{\gamma} = ( \bfcL^{uI}_{\gamma}, \bfcL^{uII}
_{\gamma})$.

To identify the optimal kernel $\bfK_{\rm p}^T$ for each operator, the following minimization problem is formulated:
\BBEQ
\label{7.3}
\!\!\!\!\!\!\!\!\!\!\!\!\!\!\!\!\!\!\!\!\!\bfK_{\gamma}^{u T*}={\rm arg}\!\min_{\!\!\!\!\!\!\!\!\!\! {{\bf K}_{\gamma}}}
\!\sum_{k=1}^N\!|| \bfcL^{uT}_{\gamma}[\lle{\bfu_k}^T\rle](\bfx)- {\bf\Gamma}^T_k(\bfx)||^2_{l_2} 
+{\cal R}(\bfK_{\gamma}).
\EEEQ
where $\mathcal{R}(\bfK_{\rm p})$ denotes a regularization functional—typically Tikhonov regularization—introduced to stabilize the ill-posed inverse problem. The kernel $\bfK_{\gamma}$ is parameterized using a basis of Bernstein polynomials, and the optimization is carried out using the Adam optimizer \cite{Kingma`B`2014}.
This approach extends and generalizes previous contributions
\cite{You`et`2024,You`et`2020,You`et`2021,You`et`2022},
by integrating physically grounded loading protocols, systematic micromechanical averaging, and data-driven inference of nonlocal constitutive behavior. As a result, it provides a versatile and robust framework for learning surrogate operators in a broad class of heterogeneous materials.

The approaches presented in \cite{{You`et`2020}, {You`et`2024}} utilize uncompressed DNS datasets (\ref{7.1}) that retain the complete microscale displacement fields for each applied BFCS $\bfb_k(\bfx)$ (\ref{2.8}) and TCCS (\ref{2.9}), leading to substantial data storage requirements. In contrast, the present study employs compressed effective datasets ${\bfcD}^{T}$, which circumvent the need for full-field DNS. Instead, these datasets leverage micromechanical averaging to extract effective properties in a more computationally efficient manner. In the case of a linearized, homogeneous peridynamic medium subjected to remote homogeneous boundary conditions—whether of type (\ref{2.10}) or (\ref{2.11})—the classical peridynamic theory directly relates the material moduli to its constitutive law \cite{Silling`et`2003}. Analogously, for a surrogate homogeneous medium under the same boundary conditions, the effective stiffness tensor can be expressed in closed form as
\BBEQ
\label{7.4}
\bfL^*&=&\int\bfK^I_{\sigma}(|\bfx-\bfy|)(\bfy-\bfx)~d\bfy,
\EEEQ
yielding a compact and interpretable representation of the homogenized elastic behavior.

The surrogate operators defined in Eqs. (\ref{7.2}) and (\ref{7.3}) are expressed in a form analogous to the peridynamic framework originally proposed by Silling \cite{Silling`2000}. A related approach was introduced by Buryachenko \cite{Buryachenko`2022}, who formulated surrogate operators based on the concept of strongly nonlocal strain-type models, as pioneered by Eringen (see e.g., \cite{Eringen`2002}).
\BBEQ
\!\!\!\!\!\!\!\!\!\!\!\!\!\!\bfcL^{\epsilon T}_{\gamma}[\lle{\bfep}^T_k\rle](\bfx) &=& {\bf \Gamma}^T(\bfx), \nonumber\\
\label{7.5}
\!\!\!\!\!\!\!\!\! \!\!\!\!\!\bfcL^{\epsilon T}_{\gamma}[\lle{\bfep}^T_k\rle](\bfx) &=&
\!\!\int \!\!\bfK^{\epsilon T}_{\rm p}(|\bfx-\bfy|) \lle{\bfep^{T}}_k\rle(\bfy)~d\bfy
\EEEQ
with notation analogous to that in Eq. (\ref{7.2}). The kernels $\bfK_{\rm p}^{\epsilon T*}$ are obtained by solving the optimization problem from Eq. (\ref{7.3}), with the replacement $\bfcL^{uT}_{\rm p}\lle{\bfu}^T_k\rle \to \bfcL^{\epsilon T}_{\rm p}\lle{\bfep}^T_k\rle$.
Under homogeneous loading conditions as in Eq. (\ref{2.10}) and constant temperature $\theta = {\rm const.}$, the effective properties are expressed through the surrogate operators
\BBEQ
\label{7.6}
\lle\bfsi\rle&=&\bfL^{*}\lle\bfep\rle+\bfal^*,\ \ \bfL^*=\int(\bfK^{\epsilon I}_{\sigma})^{\top}(|\bfx-\bfy|)~d\bfy,\\
\label{7.7}
\bfal^*&=&-\int(\bfK^{\epsilon I}_{\sigma})^{\top}(|\bfx-\bfy|)~d\bfy\bfbe^{(0)}
\nonumber\\
&-&\bar v_i^{-1}\int \int_{v_i}(\bfK^{\epsilon I}_{\lle\sigma\rle_i})^{\top}(|\bfx-\bfy|)\bfbe_1(\bfx)~d\bfx d\bfy.
\EEEQ

The surrogate operators defined in Eqs. (\ref{7.2}) and (\ref{7.3}) are inherently static and tailored for modeling linear material responses. To extend beyond this limitation, nonlocal neural operators have emerged as powerful tools capable of learning generalized mappings between function spaces, thus offering enhanced modeling flexibility and expressiveness \cite{{Lanthaler`et`2024}, {Li`et`2003}}.
Conventional architectures such as fully connected neural networks (FCNNs) are designed to represent local nonlinear operators.
Specifically, an $L$-layer FCNN $\Psi(\bfx)$: $\bfR^{\rm d_{\bf x}}\to \bfR^{\rm d_{\rm \bf u}}$ maps input $\bfx$ to output $\bfu$ through successive transformations:
\BBEQ
\label{7.8}
\!\!\!\!\!\!\!\!\!\!\!\!\!\!\!\!\!\bfz^l(\bfx)=\bfcA({\bf w}^l \bfz^{l-1}(\bfx)+\bfb^l), \ \bfu(\bfx)={\bf w}^L \bfz^{L-1}(\bfx)+\bfb^L,
\EEEQ
where $\bfcA$ denotes an activation function (e.g., ReLU or tanh), and $ ({\bf w}^l, \bfb^l)_{l=1}^L$ are the trainable parameters. However, such models are local, meaning the prediction at each point $\bfx$ relies solely on information from that same location. Nonlocal neural operators, by contrast, embed spatial dependencies directly into the architecture. They do this by integrating information over the entire domain, typically via kernel-based integral operators. A representative nonlocal layer has the structure:
\BB
\label{7.9}
\bfz^l(\bfx)= \bfcA({\bf w}^l \bfz^{l-1}(\bfx)+\bfb^l+(\bfcK^l(\bfz^{l-1})(\bfx)),
\EE
where $\bfcK^l$ is a learnable nonlocal kernel operator that aggregates contributions from surrounding points.
Several neural operator variants implement this paradigm in different ways, including Deep Operator Networks (DeepONets), PCA-Net, Graph Neural Operators, Fourier Neural Operators (FNOs), and Laplace Neural Operators (LNOs). These frameworks vary primarily in how the kernel $\bfK^l$ is constructed and applied, determining the form and extent of the nonlocal interactions. For detailed comparisons and performance evaluations across applications, refer to recent benchmark studies such as \cite{{Kumara`Y`2023}, {Lanthaler`et`2024},{Gosmani`et`2022}, {HuZ`et`2024}}

The Peridynamic Neural Operator (PNO) \cite{Jafarzadeh`et`2024} provides a powerful surrogate operator $\bfcG$—approximating ($\{\bfu(\bfb,\bfx),\bfb(\bfx)\}\in\bfcD^{\rm DNS})$
\BB
\label{7.10}
\bfcG(\bfu)(\bfx) \approx -\bfb(\bfx)
\EE
for capturing the response of highly nonlinear, anisotropic, and heterogeneous materials. Unlike classical models that rely on predefined constitutive laws (e.g., Eq. (\ref{7.2})), PNO enables data-driven modeling with improved accuracy and computational efficiency.
Physics-Informed Neural Networks (PINNs) \cite{{Raissi`et`2019}, {Karniadakis`et`2021}, {HuZ`et`2024}, {Cuomo`et`2022}, {Haghighata`et`2021}, {Harandi`et`2024}, {Kim`L`2024}, {Ren`L`2024}} enhance neural network training by embedding physical governing equations—such as Eq. (\ref{2.5})—directly into the loss function through their residuals. This ensures the learned solutions honor the fundamental physics of the system under study.
By combining PINNs with neural operator architectures \cite{{Faroughi`et`2024}, {Gosmani`et`2022}, {Wang`Y`2024}}, it becomes possible to model nonlinear material behavior, microstructural complexity, and nonlocal effects in a unified data-physics hybrid framework. These approaches have shown excellent generalization capabilities, even in high-dimensional and multiscale settings. However, a key limitation remains: such models are typically constrained to finite computational domains, which poses challenges for directly modeling infinite or unbounded media—a requirement in many micromechanics and homogenization problems.

The micromechanical model of CAM (detailed in Sections 4 and 5) can be seamlessly embedded into frameworks like PNO by substituting the conventional full-field dataset ${\bfcD}^{\rm DNS}$ with the compressed statistical effective datasets ${\bfcD}^T
=\{\bfcD^I,\bfcD^{II}\}$ (\ref{6.1}) and (\ref{6.2}).
In particular, instead of the PNO surrogate operator $\bfcG(\bfu)(\bfx) $ in (\ref{7.10}), we now consider four thermoelastic operators:
\BBEQ
\label{7.11}
\bfcG^{uT}[\lle{\bfu}^T_k\rle](\bfx) &=& {\bf \Gamma}^T(\bfx),
\EEEQ
where $T=\{I,II\}$ and ${\bf \Gamma}^T(\bfx)$ are defined in Eq. (\ref{7.2})).
This leads to the formulation of a CAM-based neural network (CAMNN) approach. By doing so, the model inherently eliminates boundary layer and size effects, which often limit the generalization of traditional neural operators to arbitrary domain geometries (as noted in \cite{Jafarzadeh`et`2024}).
Importantly, the CAMNN framework is naturally generalized to the entire space $\mathbb{R}^d$, avoiding the dependence on specific domain shapes or boundary condition residual losses that commonly appear in other recent methods (e.g., \cite{{Eghbalpoor`S`2024},{Ning`et`2023},{Yu`Z`2024a},{Yu`Z`2024b},{Zhou`Y`2024}}).
The principal advantage of using ${\bfcD}$ instead of ${\bfcD}^{\rm DNS}$ lies in the fact that ${\bfcD}$ is constructed for compact-support looading (\ref{2.8}) and (\ref{2.9}), and explicitly incorporates the RVE. This enables the definition of a nonlocal analog of the effective concentration tensor (\ref{6.4}$_2$), thereby extending the applicability of the CAMNN framework to nonlinear phenomena such as fracture and plasticity.
Crucially, preparing training data at the RVE level removes size and boundary effects at the source, which cannot be fully addressed or compensated for during or after training. Without RVE-based preparation, these artifacts persist and degrade the reliability of surrogate operators defined by Eq. (\ref{7.2}) (or (\ref{7.11})). Therefore, the RVE concept is indispensable for achieving physically meaningful and scale-invariant ML\&NN predictions in computational micromechanics.

{\color{black} Thus, when combined with BFCS (\ref{2.8}) and TCCS (\ref{2.9}) loading, the newly introduced AGIEs, the novel RVE concept, the refined effective dataset, and ML\&NN tools together form a practical predictive framework:}
\BB
\label{7.12}
{\rm BFCS}+{\rm TCCS}\ \to {\rm AGIE}\to \ \to \ {\rm RVE}\ \to\ \ \bfcD^T\ \to\ {\rm ML}\&{\rm NN}.
\EE
{\color{black}
The concepts of AGIE, the new RVE, and the effective dataset $\bfcD^T$ -- as well as their integration into ML\&NN methodology--represent fundamentally new theoretical constructs. To the best of the author’s knowledge, no prior publications have introduced these ideas in their present form. In particular, it is the combined use of these concepts that renders the proposed approach (\ref{7.12}) a foundational advance, arguably the most significant step since the original foundations of micromechanics established by Poisson, Faraday, Mossotti, Clausius, Lorenz, and Maxwell (1824–1879) (see for details \cite{Buryachenko`2025}).}

A concise schematic of the proposed methodology (\ref{7.12}) is shown in Fig. 6. Block 1 provides the input description of the BFCS (\ref{2.8}) and TCCS (\ref{2.9}). The central Block 2 is dedicated to evaluating the effective dataset $\bfcD^T$.
A key point is that these effective datasets $\bfcD^T$ contain no information about the composite microstructure—such as the inclusion concentration $c^{(1)}$, the characteristic size $a$, or the local/nonlocal constitutive properties of individual phases. Likewise, they are fully independent of the computational method employed (e.g., volume integral equation method, see Section 5, or interface integral equation method, see \cite{Buryachenko`2022}). At the same time, all finite-size effects, boundary layers, and edge artifacts are completely eliminated.
The resulting datasets $\bfcD^T$ are then passed to Block 3, the ML\&NN module. In conventional implementations, the PINN framework (and its variants) enforces governing- and boundary-condition losses \cite{Raissi`et`2019, Karniadakis`et`2021, Hu`et`2024}, thereby ensuring physically consistent solutions. In the present approach, however, this step is redundant: the effective dataset $\bfcD^T$ already encodes these constraints.
Moreover, the internal workings of Block 2 are irrelevant to Block 3, and vice versa. In this sense, Blocks 2 and 3 operate as mutual “black boxes” (indicated in black in Fig. 6). For the first time, the methodology is cast into a formalized modular framework, where each block can be developed independently without requiring in-depth knowledge of the other. In collaborative software development, the Block 2 and Block 3 teams may refine their respective modules at any stage without detailed coordination; only minimal high-level alignment is needed to ensure proper interfacing and data exchange.

\vspace{0.mm}
\hspace{2mm} \parbox{12.8cm}{
\centering \epsfig{figure=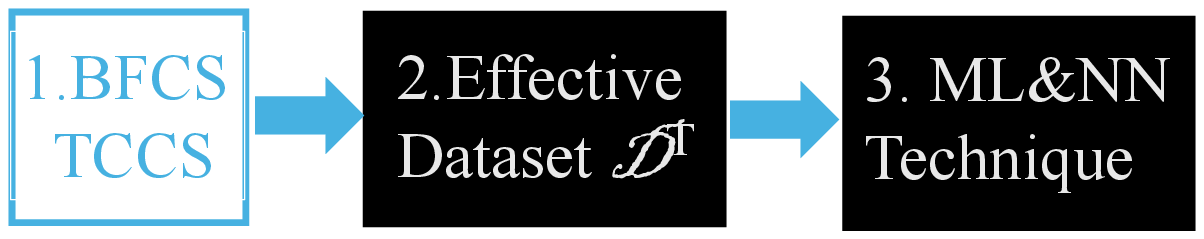, width=9.cm}\\ \vspace{-22.mm}
\vspace{-7.mm}
\vspace{-67.mm} \tenrm \baselineskip=8pt {\color{black}
{\hspace{5.mm}{\sc Fig. 6:} Scheme of proposed approach}}
}\vspace{1.mm}

{\color{black} Finally, the effective dataset $\bfcD^T$ is transferred to Block 3, the ML\&NN module. In standard practice, PINNs (and their variants) enforce physics by minimizing residuals of governing equations, boundary conditions, and initial conditions for a prescribed material model, making them essentially model-driven rather than data-driven \cite{Raissi`et`2019, Karniadakis`et`2021, Hu`et`2024}.
In contrast, within the proposed approach (see Fig. 6), explicit enforcement of governing- and boundary-condition losses becomes unnecessary in the final step $\bfcD^T$ , since these constraints are already explicitly embedded in the effective dataset $\bfcD^T$ itself. Thus, while both PINNs and the present approach are physics-informed, the physics in Fig. 6 is enforced prior to optimization, rather than during it.
As a result, the physical consistency embedded in $\bfcD^T$ is achieved at the highest attainable fidelity, with controlled numerical accuracy. Although a comparison of computational efficiency between these two strategies would be of interest, the elimination of boundary-layer effects clearly favors the present approach. Consequently, incorporating
$\bfcD^T$ into PNO and HeteroPNO by directly replacing $\bfcD^{\rm DNS} \to\bfcD^T$ is more natural than embedding $\bfcD^T$ into a PINN framework.}

Therefore, the entire micromechanical procedure ultimately converges to the construction of a single effective dataset, $\bfcD^T$. Once this dataset is obtained, the subsequent computational framework becomes fully independent of the underlying microstructure, phase properties, and of the specific numerical method (FEA, FFT, etc.) used to generate $\bfcD^T$. This effective dataset may then be efficiently approximated by ML\&NN approaches using any chosen—or even completely undefined a priori—surrogate model. No additional micromechanical simulations are needed. In effect, once the dataset is assembled, the elaborate micromechanical machinery can be switched off, and all subsequent analyses become instantaneous.

\section{ Conclusion}
\vspace{.0mm}

The term micromechanics has traditionally evoked associations with classical constructs such as effective moduli, linear local elasticity, Hill’s representative volume element (RVE) \cite{Hill`1963}, and the application of remote homogeneous boundary conditions. These well-established notions have provided a robust and systematic foundation for understanding microscale material behavior. Yet, this classical framework often imposes restrictive assumptions—an intellectual {\it Procrustean bed}—that can limit the scope of analysis and hinder innovation. In contrast, our approach moves beyond these traditional boundaries by reexamining and reconfiguring the core principles of micromechanics. This redefinition opens the door to a more versatile and expansive view of microscale interactions, accommodating complexities that classical theories overlook. By discarding rigid constraints, we introduce novel modeling strategies that more accurately reflect the diverse and often nonlocal nature of real materials.

At the core of this paradigm shift is the replacement of traditional remote homogeneous boundary conditions (\ref{2.10}) with alternatives such as BFCS (\ref{2.8}) and TCCS (\ref{2.9}). This transition eliminates many of the constraints typically associated with conventional micromechanics approaches.
By employing compact support loading conditions (\ref{2.8}) and (\ref{2.9}) as training inputs, we introduce a redefined RVE concept grounded in intrinsic phase-specific field concentration factors. This advancement effectively overcomes persistent issues like sample size dependence, boundary layer phenomena, and edge distortions.
Incorporating this generalized RVE approach into effective dataset construction naturally aligns the framework with ML\&NN methodologies for predicting nonlocal surrogate operators.
Within this setting, we introduce a comprehensive and physically motivated framework—AGIE-CAMNN—that is built for generalization. Its foundational assumptions are not rigid prescriptions but are flexible components that can be adapted, refined, or discarded to meet the evolving demands of both theoretical development and practical implementation.

Beyond the theoretical differences between AGIE and GIE, it is crucial to assess their practical significance. Traditional EFH and GIE approaches (including GIE-CAM) derive their utility from solving the governing GIE under boundary conditions (\ref{2.10}) to estimate effective moduli and field concentration factors. These methods have demonstrated consistent reliability for nearly two centuries, dating back to 1824 (see for references \cite{Buryachenko`2022}).
In contrast, the practical relevance of AGIE under compact support loading conditions (\ref{2.4}) and (\ref{2.9}) remains limited—unless enhanced by ML\&NN techniques. Without the integration of the ML\&NN component, AGIE solutions generally hold little practical utility, with the possible exception of specific cases such as the laser heating problem discussed in Subsection 3.3.
This explains why, despite the foundational simplicity of AGIE in Eq. (\ref{3.25})—which is even more elementary than the original GIE formulation in Eq. (\ref{3.28})—AGIEs have historically received limited attention.
It is only through the integrated AGIE-CAMNN framework—merging AGIE with ML\&NN methods—that meaningful and practical outcomes are realized. This unified approach allows for the prediction of a broad spectrum of predefined surrogate operators, encompassing both macroscopic effective properties and local concentration fields. Crucially, it removes dependencies on sample size, boundary conditions, and edge effects. Such capabilities represent a major advancement and highlight the transformative potential of AGIE when coupled with modern data-driven intelligence.

The construction of the effective dataset $\bfcD^T$ and its integration with the ML\&NN Block (see Fig. 6) within the proposed framework opens the door to an exceptionally wide range of applications (see, for example, the discussion at the end of Section 5). Many of these problems—regardless of their apparent complexity or specificity—can often be addressed through modest adaptations or strategic reformulations of familiar problem classes. {\color{black} In this work, the author deliberately focuses on the conceptual ideas and schematic representations of these reductions, emphasizing the key principles that allow classical methodologies to be extended.
These principles are schematically illustrated in Figs. 1–3 and 6, which elucidate the conceptual framework and logical structure of the proposed theory. In contrast, Figs. 4 and 5 present quantitative numerical results that demonstrate pronounced and systematic violations of the proposed approach, revealing fundamental limitations of traditional assumptions and highlighting regimes in which the proposed framework leads to qualitatively new behavior.
Readers who are interested in implementation details, computational aspects, or practical workflows are encouraged to explore these directions on their own.
A comprehensive discussion of such application-oriented developments, although undoubtedly interesting and promising, goes beyond the scope of the present theoretical study and is intended to be addressed in future work or in subsequent applied investigations.}

Operating within a unified analytical paradigm, AGIE-CAMNN is capable of addressing a wide range of micromechanical problems. It seamlessly generalizes to CMs with random (both statistically homogeneous and inhomogeneous), periodic, or deterministic microstructures. Moreover, it is equipped to handle both linear and nonlinear responses, including coupled and uncoupled elasticity, as well as weakly nonlocal (e.g., strain/stress gradient theories) and strongly nonlocal (e.g., strain type and displacement type, peridynamic, see classification in \cite{Maugin`2017}) constitutive behaviors.
This development of AGIE-CAMNN represents the establishment of a {\it Unified Micromechanics
Theory} for heterogeneous media, signifying a pioneering breakthrough in the
field and introducing, in essence, a {\it new philosophy of micromechanics}.

\smallskip
\noindent{\bf Acknowledgments:}

{\color{black} The author acknowledges the Reviewers for the
encouraging comments that initiated a significant correction of the manuscript.}
Permission to reproduce Figs. 5a and 5b from Silling {\it et al.} \cite{Silling`et`2024} was granted by Springer Nature (License Number 6044880317237).



\end{document}